%% file: main.tex
\documentclass{article}
\usepackage[preprint]{neurips_2024}

\usepackage[utf8]{inputenc}
\usepackage[T1]{fontenc}
\usepackage{amsmath,amssymb,amsfonts}
\usepackage{mathtools}
\usepackage{booktabs}
\usepackage{multirow}
\usepackage{graphicx}
\usepackage{xcolor}
\usepackage{hyperref}
\usepackage{cleveref}
\usepackage{subcaption}
\usepackage{tikz}
\usepackage{pgfplots}
\usepackage{pgfplotstable}
\pgfplotsset{compat=1.18}
\usetikzlibrary{positioning, arrows.meta, calc, patterns, decorations.pathreplacing, fit, backgrounds}
\usepgfplotslibrary{groupplots, colorbrewer, colormaps}
\usepackage{algorithm}
\usepackage{algorithmic}
\usepackage{enumitem}
\usepackage{microtype}
\definecolor{Dzero}{HTML}{888888}
\definecolor{Dfour}{HTML}{4682B4}
\definecolor{Afour}{HTML}{008080}
\definecolor{DAfour}{HTML}{DC8C32}
\definecolor{negcol}{HTML}{B24040}
\definecolor{poscol}{HTML}{228B22}
\colorlet{Dzerofill}{Dzero!40}
\colorlet{Dzerodraw}{Dzero!60}
\colorlet{Dfourfill}{Dfour!55}
\colorlet{Dfourdraw}{Dfour!75}
\colorlet{Afourfill}{Afour!50}
\colorlet{Afourdraw}{Afour!70}
\colorlet{DAfourfill}{DAfour!50}
\colorlet{DAfourdraw}{DAfour!70}
\colorlet{Dfourdesat}{Dfour!20}
\colorlet{Dfourdesatdraw}{Dfour!40}
\colorlet{Afourdesat}{Afour!20}
\colorlet{Afourdesatdraw}{Afour!40}
\colorlet{DAfourdesat}{DAfour!20}
\colorlet{DAfourdesatdraw}{DAfour!40}

\newcommand{\R}{\mathbb{R}}

\newcommand{\atm}{a{\to}m}
\newcommand{\mta}{m{\to}a}
\newcommand{\pp}{\,\text{pp}}

\DeclareMathOperator{\ReLU}{ReLU}

\DeclareMathOperator{\softmax}{softmax}
\DeclareMathOperator{\LayerNorm}{LayerNorm}

\DeclareMathOperator{\CKA}{CKA}

\title{Descriptor-Injected Cross-Modal Learning:\\
A Systematic Exploration of Audio--MIDI Alignment\\
via Spectral and Melodic Features}

\author{
  Mariano Fern\'{a}ndez M\'{e}ndez \\
  Asociaci\'{o}n Civil AlterMundi \\
  C\'{o}rdoba, Argentina \\
  \texttt{mariano@altermundi.net}
}

\begin{document}
\maketitle

\begin{abstract}

Cross-modal retrieval between audio recordings and symbolic music representations (MIDI) remains challenging due to the fundamental modality gap between continuous waveforms and discrete event sequences.
We present a systematic exploration of \emph{descriptor injection}---the practice of augmenting modality-specific encoders with hand-crafted domain features---as a mechanism for bridging this gap.
Through a three-phase experimental campaign spanning 13 descriptor$\times$mechanism combinations, 6 architectural families, and 3 training schedules, we identify a configuration (\textsc{d4a4}) that reaches a mean $S{=}84.0\% \pm 2.7\pp$ across five independent training seeds, a $+8.8$ percentage point improvement over the descriptor-free baseline.
Causal ablation reveals that the audio-side descriptor (A4, octave-band energy dynamics) is entirely responsible for this improvement in the top dual models (\textsc{d4a4}, \textsc{d4-a4r}): zeroing A4 collapses performance by $75$--$78\pp$, while ablating the MIDI-side descriptor (D4, local pitch intervals) changes S by only $-0.4$ to $+0.6\pp$ despite improving training dynamics.
We propose \emph{reverse cross-attention}, a mechanism where descriptor tokens serve as queries attending to encoder features, achieving 163$\times$ fewer attention operations than the standard formulation while remaining competitive.
Centered Kernel Alignment (CKA) analysis reveals that descriptors double the representational alignment between audio and MIDI transformer layers ($0.435 \to 0.794$, $+82\%$), suggesting that the mechanism of improvement is cross-encoder representational convergence rather than simple feature injection.
Perturbation analysis shows that high-frequency octave bands (750--12000\,Hz) carry the majority of the discriminative signal, and that this information is encoded non-linearly ($|r| < 0.05$ correlations with output).
All experiments use the MAESTRO v3.0.0 piano dataset with a structured evaluation protocol controlling for composer and piece similarity.

\end{abstract}

\section{Introduction}
\label{sec:introduction}

Music exists simultaneously in multiple modalities: as audio waveforms captured by microphones, as symbolic sequences encoded in MIDI files, as visual score images, and as cognitive representations in the listener's mind. The ability to align representations across these modalities---particularly between audio and symbolic music---has direct applications in music retrieval, score following, automatic transcription evaluation, and musicological analysis.

The fundamental challenge of audio--MIDI alignment lies in the \emph{modality gap}: audio is a continuous-time signal shaped by instrument timbre, room acoustics, performer expression, and recording conditions, while MIDI is a discrete sequence of pitch, velocity, and duration events that abstracts away all acoustic detail. Traditional approaches bridge this gap through hand-crafted intermediate representations---chromagrams~\citep{ellis2007chromaprint, ewert2012chroma}, audio fingerprints~\citep{wang2003shazam}, or dynamic time warping on spectral features~\citep{hu2003dtw}---but these require extensive domain engineering and are brittle to variations in performance style and recording quality.

Recent advances in self-supervised audio representation learning~\citep{baevski2020wav2vec, li2024mert} and cross-modal contrastive objectives~\citep{bardes2022vicreg, radford2021clip, wu2023clap} offer a principled alternative: learn a shared embedding space where matched audio--MIDI pairs are close and unmatched pairs are distant. However, na\"{\i}ve application of contrastive learning to audio--MIDI pairs faces a bootstrapping problem: the modality gap is so large that early training gradients are noisy and convergence is slow.

Beyond these practical applications, the question of whether audio and symbolic music share discoverable structural invariants is of broader interest. Frequency-based descriptors---capturing spectral dynamics, harmonic interval patterns, or energy distribution across octave bands---offer a tractable way to test whether certain relational features can serve as cross-modal bridges. If such descriptors prove effective, this would suggest that some aspects of musical information remain legible across modalities through frequency-structured relations rather than through learned abstractions alone.

\paragraph{Hypothesis.} We hypothesize that \emph{domain-specific descriptors}---hand-crafted features capturing known invariances within each modality---can serve as a bridge during cross-modal learning. Specifically, spectral dynamics descriptors for audio and melodic interval descriptors for MIDI may provide the encoder with structured prior information that accelerates the discovery of cross-modal correspondences.

\paragraph{Approach.} We conduct a systematic three-phase exploration:
\begin{enumerate}[nosep]
  \item \textbf{Screening} (5-epoch short-horizon): 13 descriptor$\times$injection mechanism combinations, plus 11 alternative architectures across 3 families (FiLM, Mixture-of-Experts, Third Tower).
  \item \textbf{Confirmation} (30--60 epochs): The top candidates are trained to convergence with optimized learning rate schedules.
  \item \textbf{Scientific validation} (Gate~5B): A battery of 8 tests---causal ablation, transposition invariance, CKA/RSA alignment, perturbation sensitivity, linear probing, and embedding visualization---applied to the 4 canonical models.
\end{enumerate}

\paragraph{Contributions.}
\begin{enumerate}[nosep]
  \item A comprehensive taxonomy of 13 descriptor$\times$mechanism combinations for cross-modal music learning, with systematic screening results (\cref{sec:selection}).
  \item The \emph{reverse cross-attention} mechanism, where descriptor tokens serve as queries attending to encoder features, achieving $163\times$ fewer attention operations than standard cross-attention while delivering competitive performance (\cref{sec:injection}).
  \item Causal analysis demonstrating that the audio-side descriptor (A4) drives the retrieval gains, while the MIDI-side descriptor (D4) acts mainly as a training regularizer and has only weak inference-time effect in top dual models (\cref{sec:ablation}).
  \item CKA analysis revealing that descriptors double cross-encoder representational alignment, suggesting the mechanism is representational convergence rather than feature concatenation (\cref{sec:cka}).
\end{enumerate}

\section{Related Work}
\label{sec:related}

\subsection{Audio--MIDI Matching}

Classical audio--MIDI alignment relies on intermediate representations that abstract away modality-specific details. Chromagram-based methods~\citep{ellis2007chromaprint, muller2015chromabased} project both modalities into a shared pitch-class energy space, then apply DTW~\citep{hu2003dtw} or subsequence matching. Audio fingerprinting approaches, pioneered by Shazam~\citep{wang2003shazam}, extract sparse time-frequency landmarks and match via hash voting, achieving real-time performance but requiring explicit feature engineering. Score-informed source separation~\citep{ewert2012chroma} uses MIDI as a guide for audio decomposition, demonstrating that symbolic information can meaningfully constrain audio processing.

\subsection{Self-Supervised Audio Representations}

The success of BERT-style pre-training~\citep{devlin2019bert} has inspired self-supervised approaches for audio. wav2vec~2.0~\citep{baevski2020wav2vec} learns speech representations via masked prediction over quantized latent features. MERT~\citep{li2024mert} extends this paradigm to music, pre-training on 160K hours of music audio and achieving strong performance on downstream MIR tasks. CLAP~\citep{wu2023clap} and MuLan~\citep{huang2022mulan} learn joint audio-language spaces via contrastive objectives, demonstrating that cross-modal alignment can emerge from large-scale training.

\subsection{Cross-Modal Contrastive Learning}

VICReg~\citep{bardes2022vicreg} regularizes representation learning by jointly optimizing \emph{variance} (preventing collapse), \emph{invariance} (aligning matched pairs), and \emph{covariance} (decorrelating dimensions). Unlike InfoNCE-based methods~\citep{he2020moco, chen2020simclr}, VICReg does not require negative pairs or large batch sizes, making it suitable for settings with limited data diversity. Barlow Twins~\citep{zbontar2021barlow} achieves similar goals by minimizing cross-correlation redundancy. CLIP~\citep{radford2021clip} scales contrastive learning to 400M image-text pairs, demonstrating that cross-modal alignment can produce powerful zero-shot representations.

\subsection{Feature Conditioning in Neural Networks}

Feature-wise Linear Modulation (FiLM)~\citep{perez2018film} conditions network activations via learned affine transformations $\gamma \odot F + \beta$, originally proposed for visual reasoning. Mixture-of-Experts (MoE)~\citep{shazeer2017moe} routes inputs through specialized sub-networks via learned gating. Cross-attention~\citep{vaswani2017attention, jaegle2021perceiver} allows one set of representations to attend to another, enabling flexible information flow between modalities or feature streams.

\subsection{Representational Similarity Analysis}

Centered Kernel Alignment (CKA)~\citep{kornblith2019cka} provides a principled measure of representational similarity between neural network layers, invariant to orthogonal transformations and isotropic scaling. Representational Similarity Analysis (RSA)~\citep{kriegeskorte2008rsa} compares representations via pairwise distance correlations. Linear probing~\citep{alain2017probing} evaluates the linear decodability of specific features from frozen representations, providing a complementary view to geometric similarity measures.

\section{Method}
\label{sec:method}

\subsection{Problem Formulation}
\label{sec:formulation}

Given an audio segment $a \in \R^{T_a}$ sampled at $f_s = 24$\,kHz ($T_a = 96{,}000$ for 4-second segments) and a corresponding MIDI segment $m = \{(p_i, v_i, d_i, o_i)\}_{i=1}^{N}$ where $p_i \in \{0, \ldots, 127\}$ is pitch, $v_i \in \{0, \ldots, 127\}$ is velocity, $d_i$ is duration, and $o_i$ is onset time, we seek to learn encoders $f_a: \R^{T_a} \to \R^D$ and $f_m: \mathcal{M} \to \R^D$ mapping both modalities into a shared $D$-dimensional embedding space (with $D = 256$) such that
\begin{equation}
  \text{sim}(f_a(a), f_m(m)) \gg \text{sim}(f_a(a), f_m(m'))
  \quad \text{for matched } (a, m) \text{ and unmatched } m' \neq m.
\end{equation}

We define the primary evaluation metric as
\begin{equation}
  S = \min\!\big(\text{R@10}_{\atm},\; \text{R@10}_{\mta}\big),
  \label{eq:S}
\end{equation}
where $\text{R@10}_{\atm}$ is Recall@10 for audio-to-MIDI retrieval and $\text{R@10}_{\mta}$ is the reverse. The $\min$ ensures that both retrieval directions are simultaneously effective.

\subsection{Base Architecture}
\label{sec:architecture}

Our architecture consists of modality-specific encoders followed by projection heads that map to the shared embedding space (\cref{fig:architecture}). Full encoder specifications and parameter breakdowns are provided in \cref{app:architecture}.

\input{figures/architecture}

\paragraph{Audio encoder (MERTEncoderLite).}
The audio encoder follows the MERTEncoderLite architecture~\citep{li2024mert}: a 4-layer 1D CNN with GroupNorm and GELU activations ($1 \to 512 \to 512 \to 512 \to 1024$ channels, $40\times$ total downsampling) produces $h \in \R^{B \times 2400 \times 1024}$ frame features, which are summed with learnable positional embeddings and processed by a 4-layer Transformer encoder ($d_{\text{model}}{=}1024$, $n_{\text{heads}}{=}8$, $d_{\text{ff}}{=}4096$, GELU, dropout $0.1$). The final representation $z_a \in \R^{B \times 1024}$ is obtained via mean pooling over the Transformer output.

\paragraph{MIDI encoder.}
The MIDI encoder processes discrete event sequences through learned embeddings and a pre-norm Transformer. Each note event's pitch ($\R^{256}$), velocity ($\R^{128}$), and duration ($\R^{128}$) are independently embedded, concatenated to form a 512-dimensional vector, linearly projected, and LayerNorm-ed. Sinusoidal positional encodings are added, and the sequence is processed by a 4-layer pre-norm Transformer ($d_{\text{model}}{=}512$, $n_{\text{heads}}{=}8$, $d_{\text{ff}}{=}2048$, GELU, dropout $0.1$). The output $z_m \in \R^{B \times 512}$ is obtained via masked mean pooling over valid (non-padding) positions, followed by a final LayerNorm.

\paragraph{Projection heads.}
Both encoders are followed by 3-layer MLP projection heads with BatchNorm and ReLU after each hidden layer, mapping to the shared 256-dimensional space. The audio head has dimensions $1024 \to 512 \to 512 \to 256$; the MIDI head is $512 \to 512 \to 512 \to 256$.

\paragraph{Parameter counts.} The base model (D0, no descriptors) contains approximately 74.2M parameters: ${\sim}60$M in the audio encoder, ${\sim}13$M in the MIDI encoder, and ${\sim}1.2$M in both projection heads.

\subsection{VICReg Loss}
\label{sec:vicreg}

We train with VICReg~\citep{bardes2022vicreg}, which jointly optimizes three terms over a batch of $B$ matched pairs $\{(z_a^{(i)}, z_m^{(i)})\}_{i=1}^B$ projected to embeddings $\hat{z}_a = g_a(z_a)$ and $\hat{z}_m = g_m(z_m)$:

\begin{equation}
  \mathcal{L}_{\text{VICReg}} = \lambda_{\text{inv}}\,\mathcal{L}_{\text{inv}} + \lambda_{\text{var}}\,\mathcal{L}_{\text{var}} + \lambda_{\text{cov}}\,\mathcal{L}_{\text{cov}},
  \label{eq:vicreg}
\end{equation}

The invariance term $\mathcal{L}_{\text{inv}}$ minimizes the MSE between matched embedding pairs; the variance term $\mathcal{L}_{\text{var}}$ applies a hinge loss on the per-dimension standard deviation to prevent representational collapse; and the covariance term $\mathcal{L}_{\text{cov}}$ penalizes off-diagonal elements of the embedding covariance matrix to decorrelate dimensions \citep[see][for full definitions]{bardes2022vicreg}. We use $\lambda_{\text{inv}} = 10$, $\lambda_{\text{var}} = 10$, $\lambda_{\text{cov}} = 1$ throughout all experiments.

\subsection{Descriptors}
\label{sec:descriptors}

We design two domain-specific descriptors for the modalities represented in Gate~5B. Additional descriptors explored during screening are detailed in \cref{app:descriptors}.

\subsubsection{A4: Octave-Band Energy Dynamics (Audio)}
\label{sec:a4}

The A4 descriptor captures temporal dynamics of spectral energy across logarithmically-spaced frequency bands. Given an audio waveform, we compute:

\begin{enumerate}[nosep]
  \item \textbf{STFT}: Compute the magnitude spectrogram with $n_{\text{FFT}} = 2048$, hop length $= 512$, yielding frequency resolution $\Delta f \approx 11.72$\,Hz/bin at $f_s = 24$\,kHz. The output has $T_{\text{stft}} = 188$ frames for a 4-second segment.

  \item \textbf{Log-magnitude}: $M[k, t] = \log(1 + |X[k, t]|)$, where $X$ is the STFT.

  \item \textbf{Octave-band aggregation}: We define 8 bands with logarithmically-spaced boundaries (\cref{tab:bands}), and average the log-magnitude within each band:
  \begin{equation}
    \bar{M}[b, t] = \frac{1}{k_b^{\text{end}} - k_b^{\text{start}}} \sum_{k=k_b^{\text{start}}}^{k_b^{\text{end}}-1} M[k, t].
  \end{equation}

  \item \textbf{Temporal delta}: $\Delta M[b, t] = \bar{M}[b, t+1] - \bar{M}[b, t]$.

  \item \textbf{Normalization}: Per-band z-score normalization within each sample:
  \begin{equation}
    A4[b, t] = \frac{\Delta M[b, t] - \mu_b}{\sigma_b + \epsilon}, \quad \mu_b = \frac{1}{T}\sum_t \Delta M[b,t], \quad \sigma_b = \sqrt{\frac{1}{T}\sum_t (\Delta M[b,t] - \mu_b)^2}.
    \label{eq:a4}
  \end{equation}
\end{enumerate}

The resulting descriptor $A4 \in \R^{B \times 188 \times 8}$ captures how spectral energy \emph{changes} over time in each frequency band, providing the audio encoder with explicit temporal dynamics information.

\begin{table}[t]
  \centering
  \caption{A4 octave-band definitions. STFT bins correspond to $n_{\text{FFT}} = 2048$ at $f_s = 24$\,kHz.}
  \label{tab:bands}
  \begin{tabular}{clccc}
    \toprule
    Band & Frequency range & STFT bins & Width (Hz) & Musical content \\
    \midrule
    0 & 47\,--\,94\,Hz   & 4\,--\,8    & 47    & Bass fundamentals \\
    1 & 94\,--\,188\,Hz  & 8\,--\,16   & 94    & Bass harmonics \\
    2 & 188\,--\,375\,Hz & 16\,--\,32  & 187   & Low-mid register \\
    3 & 375\,--\,750\,Hz & 32\,--\,64  & 375   & Mid register \\
    4 & 750\,--\,1500\,Hz & 64\,--\,128 & 750   & Upper-mid/presence \\
    5 & 1500\,--\,3000\,Hz & 128\,--\,256 & 1500 & Presence/harmonics \\
    6 & 3000\,--\,6000\,Hz & 256\,--\,512 & 3000 & Brilliance \\
    7 & 6000\,--\,12000\,Hz & 512\,--\,1025 & 6000 & Air/high harmonics \\
    \bottomrule
  \end{tabular}
\end{table}

\subsubsection{D4: Local Pitch Intervals (MIDI)}
\label{sec:d4}

The D4 descriptor encodes local melodic context for each MIDI note using 4 features derived from pitch differences with neighboring notes:
\begin{equation}
  D4_i = \begin{bmatrix}
    (p_i - p_{i-1}) / 24 \\[2pt]
    (p_{i+1} - p_i) / 24 \\[2pt]
    \text{clamp}\!\big((p_i - p_{i-1})/12,\; -2,\; 2\big) / 2 \\[2pt]
    \text{clamp}\!\big((p_{i+1} - p_i)/12,\; -2,\; 2\big) / 2
  \end{bmatrix} \in \R^4,
  \label{eq:d4}
\end{equation}
where $p_i$ is the MIDI pitch (0--127). The first two features capture raw semitone intervals normalized by two octaves, while the latter two apply clamping to emphasize intervals within $\pm 2$ octaves. Boundary notes ($i=1$ or $i=N$) use zero-padding.

The output $D4 \in \R^{B \times N \times 4}$ provides the MIDI encoder with explicit intervallic context, analogous to how music theory emphasizes relative pitch over absolute pitch.

\subsection{Injection Mechanisms}
\label{sec:injection}

We explore four mechanisms for injecting descriptors into encoders.

\subsubsection{Concatenation}
\label{sec:concat}

The simplest approach concatenates the descriptor with encoder-internal features and projects back to the original dimension:
\begin{equation}
  h'_t = \LayerNorm\!\big(W_{\text{proj}}[h_t \,\|\, d_t]\big),
  \label{eq:concat}
\end{equation}
where $h_t \in \R^{d_{\text{model}}}$ is the encoder feature at time $t$, $d_t$ is the descriptor value, and $W_{\text{proj}} \in \R^{d_{\text{model}} \times (d_{\text{model}} + d_{\text{desc}})}$.

For A4 injection into the audio encoder: $h_t \in \R^{1024}$, $d_t \in \R^{8}$, so $W_{\text{proj}} \in \R^{1024 \times 1032}$. The concatenation occurs after the CNN and before the Transformer.

For D4 injection into the MIDI encoder: $e_t \in \R^{512}$, $d_t \in \R^{4}$, so $W_{\text{proj}} \in \R^{512 \times 516}$. The concatenation occurs after event embedding and before the Transformer.

\subsubsection{Standard Cross-Attention}
\label{sec:standard_xatt}

In standard cross-attention, encoder features serve as queries attending to descriptor features:
\begin{equation}
  \text{XAttn}(F, D) = \softmax\!\left(\frac{Q_F \, K_D^\top}{\sqrt{d_k}}\right) V_D,
  \label{eq:standard_xatt}
\end{equation}
where $Q_F = F\,W_Q \in \R^{T_F \times d_k}$, $K_D = D_{\text{proj}}\,W_K \in \R^{T_D \times d_k}$, $V_D = D_{\text{proj}}\,W_V \in \R^{T_D \times d_v}$. Here $T_F = 2400$ (CNN output frames) and $T_D$ depends on the descriptor.

The attention complexity is $O(T_F \cdot T_D \cdot d_k)$, dominated by the $T_F = 2400$ query tokens.

\subsubsection{Reverse Cross-Attention (Key Contribution)}
\label{sec:reverse_xatt}

We propose \emph{reverse cross-attention}, where \textbf{descriptor tokens serve as queries} attending to encoder features:
\begin{equation}
  \text{RevXAttn}(D, F) = \softmax\!\left(\frac{Q_D \, K_F^\top}{\sqrt{d_k}}\right) V_F,
  \label{eq:reverse_xatt}
\end{equation}
where $Q_D = D_{\text{proj}}\,W_Q \in \R^{T_D \times d_k}$, $K_F = F\,W_K \in \R^{T_F \times d_k}$, $V_F = F\,W_V \in \R^{T_F \times d_v}$.

For A4 reverse cross-attention: $T_D = 188$ (native STFT frames) and $T_F = 2400$ (CNN output). The descriptor is projected to $d_{\text{model}} = 1024$ and augmented with learnable positional embeddings before serving as queries.

\paragraph{Complexity advantage.} The critical difference is in the \emph{output} sequence length: reverse cross-attention produces $T_D = 188$ tokens (which are then mean-pooled), while standard cross-attention produces $T_F = 2400$ tokens. Since subsequent Transformer self-attention has $O(T^2)$ complexity, processing 188 vs.\ 2400 tokens yields:
\begin{equation}
  \text{Speedup} = \frac{T_F^2}{T_D^2} = \frac{2400^2}{188^2} \approx 163\times.
  \label{eq:speedup}
\end{equation}

This is conceptually related to the Perceiver architecture~\citep{jaegle2021perceiver}, which uses a small set of latent tokens as queries to compress high-dimensional inputs. In our case, the ``latent'' tokens are not learned from scratch but carry structured domain knowledge via the descriptor.

\subsubsection{Combined Mechanisms}

Two combined configurations are evaluated in Gate~5B:
\begin{itemize}[nosep]
  \item \textbf{d4a4}: D4 concatenation in MIDI encoder + A4 concatenation in audio encoder.
  \item \textbf{d4-a4r}: D4 concatenation in MIDI encoder + A4 reverse cross-attention in audio encoder.
\end{itemize}

\subsection{Training Protocol}
\label{sec:training}

\paragraph{Dataset.} We use MAESTRO v3.0.0~\citep{hawthorne2019maestro}, containing 1,276 classical piano performances with aligned audio and MIDI (${\sim}200$ hours). Audio--MIDI pairs are segmented into 4-second windows with 1-second hop, yielding ${\sim}13{,}000$ training segments at $f_s = 24$\,kHz.

\paragraph{Optimizer.} AdamW~\citep{loshchilov2019adamw} with weight decay $0.01$, $\beta = (0.9, 0.999)$, and batch size 16.

\paragraph{Learning rate schedule.} We employ a cosine-with-tail schedule~\citep{loshchilov2017sgdr}: linear warmup for 500 steps, cosine annealing with reference period of 30 epochs, a floor of $10\%$ of base LR, and a linear tail decaying to $2\%$ in the final epoch.

\paragraph{Hardware.} Primary experiments on a single NVIDIA RTX~3090 (24\,GB). Multi-seed validation and parameter-matched controls on a cluster of 36$\times$ NVIDIA A30 (24\,GB each) via SLURM. Full hyperparameters and scheduler details are provided in \cref{app:hyperparams}.

\section{Descriptor and Mechanism Selection}
\label{sec:selection}

\subsection{Short-Horizon Screening (5 Epochs, 13 Arms)}
\label{sec:screening}

We conduct a systematic screening of 13 descriptor$\times$mechanism combinations over 5 training epochs. Each arm uses identical hyperparameters except for the descriptor and injection mechanism. Results are shown in \cref{tab:screening} and \cref{fig:screening}.

\input{figures/screening}

\begin{table}[t]
  \centering
  \caption{Gate~4.3 screening results (5 epochs). S = min(R@10$_{\atm}$, R@10$_{\mta}$). $\Delta$ is relative to the D0 baseline. Arms above the horizontal line outperform D0.}
  \label{tab:screening}
  \begin{tabular}{clllccc}
    \toprule
    Rank & Arm & Descriptor & Mechanism & S (\%) & Hard neg (\%) & $\Delta$ (pp) \\
    \midrule
    1 & \textbf{d4a4}  & D4+A4 & Dual concat        & \textbf{69.8} & 91.6 & $+9.6$ \\
    2 & A4r            & A4    & Reverse cross-att   & 68.6          & 91.6 & $+8.4$ \\
    3 & D4r            & D4    & Reverse cross-att   & 64.2          & 93.2 & $+4.0$ \\
    4 & D4             & D4    & Concat              & 63.6          & 91.2 & $+3.4$ \\
    4 & A4             & A4    & Concat              & 63.6          & 92.4 & $+3.4$ \\
    6 & A4x            & A4    & Cross-attention     & 62.6          & 92.4 & $+2.4$ \\
    7 & A7x            & A7    & Cross-attention     & 62.2          & 92.0 & $+2.0$ \\
    \midrule
    8 & \textbf{D0}    & ---   & \emph{Baseline}     & \textbf{60.2} & 90.0 & --- \\
    \midrule
    9 & D4x            & D4    & Cross-attention     & 60.0          & 91.4 & $-0.2$ \\
    10 & A7             & A7    & Concat              & 58.8          & 90.2 & $-1.4$ \\
    10 & A9             & A9    & Concat              & 58.8          & 90.4 & $-1.4$ \\
    12 & A8             & A8    & Concat              & 57.4          & 90.6 & $-2.8$ \\
    13 & d4a4cm         & D4+A4 & Cross-modal         & 52.4          & 89.6 & $-7.8$ \\
    \bottomrule
  \end{tabular}
\end{table}

\paragraph{Key observations.}
\begin{enumerate}[nosep]
  \item \textbf{Same-modality injection dominates}: The top 7 arms all inject descriptors into their own modality's encoder. The single cross-modal injection (d4a4cm, injecting audio features into MIDI and vice versa) is the worst arm ($-7.8\pp$).
  \item \textbf{Dual injection is best}: d4a4 ($+9.6\pp$) combines D4 in MIDI and A4 in audio, exploiting both modalities.
  \item \textbf{Reverse cross-attention is competitive}: A4r ($+8.4\pp$) is the second-best arm despite processing only 188 tokens vs.\ 2400.
  \item \textbf{Complex descriptors underperform}: A7 (rational attractor, 12\,dim), A8 (onset-chroma, 12\,dim), and A9 (IDF-attractor, 12\,dim) all perform at or below baseline, while the simpler A4 (8\,dim) and D4 (4\,dim) excel.
\end{enumerate}

\subsection{Architecture Families (Gate 4.4, 11 Arms)}
\label{sec:arch_families}

We additionally explore three alternative architectural families for descriptor injection (\cref{tab:arch_families}).

\begin{table}[t]
  \centering
  \caption{Gate~4.4 architecture family screening (5 epochs, from d4a4 foundation). Only Third Tower with weighted loss outperforms the D0 baseline substantially.}
  \label{tab:arch_families}
  \begin{tabular}{clcccc}
    \toprule
    Rank & Arm & Family & S (\%) & $\Delta$ vs D0 (pp) \\
    \midrule
    1 & t3-wt      & Third Tower  & 67.6 & $+7.4$ \\
    2 & t3-tri     & Third Tower  & 65.0 & $+4.8$ \\
    3 & moe-a4-v2  & MoE          & 60.2 & $\pm 0.0$ \\
    4 & film-dual  & FiLM         & 59.4 & $-0.8$ \\
    4 & moe-a4-v4  & MoE          & 59.4 & $-0.8$ \\
    6 & film-a4    & FiLM         & 59.2 & $-1.0$ \\
    6 & moe-dual   & MoE          & 59.2 & $-1.0$ \\
    6 & moe-a4-v3  & MoE          & 59.2 & $-1.0$ \\
    9 & film-d4    & FiLM         & 58.6 & $-1.6$ \\
    10 & moe-a4    & MoE          & 58.2 & $-2.0$ \\
    11 & t3-anc    & Third Tower  & 42.2 & $-18.0$ \\
    \bottomrule
  \end{tabular}
\end{table}

\textbf{FiLM} ($\gamma, \beta$ modulation of Transformer features): No variant improves over the baseline, suggesting that affine transformation is too constrained for the cross-modal information flow required.

\textbf{Mixture-of-Experts} (learned routing across 4--8 experts): All variants perform near or below D0, indicating that the routing mechanism does not discover useful specialization patterns for this task.

\textbf{Third Tower} (separate descriptor encoder with auxiliary VICReg loss): The weighted-loss variant (t3-wt) achieves $S = 67.6\%$ ($+7.4\pp$), competitive with direct injection, but the anchor-based variant (t3-anc) collapses ($-18.0\pp$). Per-family Gate~4.4 results are reported in \cref{app:gate44}.

\subsection{Long-Horizon Confirmation (30--60 Epochs)}
\label{sec:long_horizon}

The top candidates from screening are trained for 30--60 epochs with their selected schedules (Cosine-30/60, Cosine-tail, Trapezoidal hold). Results (\cref{tab:long_horizon}) confirm the screening rankings in the canonical seed-42 run, with one notable result: d4a4 reaches $S = 83.8\%$ at epoch~50 under 60-epoch cosine-stretched scheduling.

\begin{table}[t]
  \centering
  \caption{Long-horizon results (30--60 epochs) from the seed-42 training run. $\Delta$: improvement over D0 at convergence. Multi-seed training replication is summarized in \cref{tab:multiseed}.}
  \label{tab:long_horizon}
  \begin{tabular}{lcccccc}
    \toprule
    Arm & Schedule & Epochs & Best S (\%) & Best epoch & Params (M) & $\Delta$ (pp) \\
    \midrule
    \textbf{d4a4}   & Cosine-60    & 60 & \textbf{83.8} & 50 & 75.5 & $+10.4$ \\
    a4r              & Cosine-30    & 30 & 82.0          & 29 & 78.6 & $+8.6$ \\
    t3-wt            & Trapez-50    & 50 & 81.2          & 50 & ---  & $+7.8$ \\
    a4r              & Cosine-tail  & 60 & 80.6          & 60 & 78.6 & $+7.2$ \\
    d4-a4r           & Cosine-30    & 30 & 79.8          & 30 & 78.9 & $+6.4$ \\
    D0               & Cosine-tail  & 60 & 73.4          & 50 & 74.2 & --- \\
    \bottomrule
  \end{tabular}
\end{table}

\section{Scientific Validation --- Gate 5B}
\label{sec:validation}

We validate the top 4 models (D0, d4a4, a4r, d4-a4r) through a battery of 8 complementary tests, each probing a different aspect of the learned representations. Unless otherwise noted, the detailed diagnostic tests below are computed on the canonical seed-42 checkpoint. The final subsection reports the five-seed training replication used for the paper's headline claims.

\subsection{Canonical Evaluation (Test 12: Scoreboard)}
\label{sec:scoreboard}

The canonical evaluation uses a structured pool protocol: 256 pool items, 500 queries, 64 hard negatives (same piece, different time), 32 semi-hard negatives (same composer, different piece), seed 42. Results are shown in \cref{tab:scoreboard}.

\begin{table}[t]
  \centering
  \caption{Canonical evaluation results (Test~12) from the seed-42 checkpoint. Pool size 256, 500 queries, seed 42. Multi-seed training replication is summarized in \cref{tab:multiseed}.}
  \label{tab:scoreboard}
  \small
  \setlength{\tabcolsep}{5pt}
  \begin{tabular}{@{}lccccccc@{}}
    \toprule
    Arm & S (\%) & R@1$_{\atm}$ & R@10$_{\atm}$ & MRR$_{\atm}$ & R@10$_{\mta}$ & MRR$_{\mta}$ & Hard neg (\%) \\
    \midrule
    D0      & 73.4 & 22.2 & 74.8 & 0.380 & 73.4 & 0.372 & 98.4 \\
    \textbf{d4a4} & \textbf{83.8} & \textbf{28.0} & \textbf{84.4} & \textbf{0.458} & \textbf{83.8} & \textbf{0.442} & \textbf{99.0} \\
    a4r     & 82.0 & 27.6 & 82.6 & 0.438 & 82.0 & 0.424 & 98.8 \\
    d4-a4r  & 79.8 & 25.2 & 81.4 & 0.426 & 79.8 & 0.420 & 99.4 \\
    \bottomrule
  \end{tabular}
\end{table}

All descriptor-augmented models substantially outperform the D0 baseline: d4a4 by $+10.4\pp$, a4r by $+8.6\pp$, and d4-a4r by $+6.4\pp$. The improvements are consistent across all metrics (R@1, R@10, MRR).

\subsection{Causal Ablation (Test 01)}
\label{sec:ablation}

To determine whether descriptors are \emph{causally} necessary at inference time (as opposed to merely useful during training), we ablate each descriptor independently using three perturbation modes: \textbf{zero} (replace descriptor with zeros), \textbf{noise} (replace with Gaussian noise matching descriptor statistics), and \textbf{shuffle} (randomly permute descriptors across the batch).

Results (\cref{tab:ablation}, \cref{fig:ablation}) reveal a stark asymmetry:

\input{figures/ablation}

\begin{table}[t]
  \centering
  \caption{Causal ablation results (Test~01). $\Delta$: change in S relative to normal inference. \textbf{A4 is causally essential; D4 has weak effect in this model family.}}
  \label{tab:ablation}
  \begin{tabular}{llccc}
    \toprule
    & Ablation mode & d4a4 ($\Delta$) & a4r ($\Delta$) & d4-a4r ($\Delta$) \\
    \midrule
    \multirow{3}{*}{\textbf{Audio (A4)}}
    & Zero    & $-76.0\pp$ & $-77.6\pp$ & $-75.4\pp$ \\
    & Noise   & $-61.8\pp$ & $-53.0\pp$ & $-53.0\pp$ \\
    & Shuffle & $-37.2\pp$ & $-32.2\pp$ & $-32.4\pp$ \\
    \midrule
    \multirow{3}{*}{MIDI (D4)}
    & Zero    & $+0.6\pp$ & --- & $-0.4\pp$ \\
    & Noise   & $+0.6\pp$ & --- & $\pm 0.0\pp$ \\
    & Shuffle & $\pm 0.0\pp$ & --- & $\pm 0.0\pp$ \\
    \bottomrule
  \end{tabular}
\end{table}

\paragraph{A4 is completely causal.} Zeroing A4 collapses S from $83.8\% \to 7.8\%$ (d4a4), $82.0\% \to 4.4\%$ (a4r), and $79.8\% \to 4.4\%$ (d4-a4r)---drops of 75--78$\pp$ that bring performance below random chance. Even the milder shuffle ablation causes drops of 32--37$\pp$. The model has become \emph{entirely dependent} on A4 for cross-modal alignment.

\paragraph{D4 has weak inference effect in this family.} In the dual top models, zeroing D4 in d4a4 produces $S = 84.4\%$ (actually $+0.6\pp$ higher than normal), and in d4-a4r produces $S = 79.4\%$ ($-0.4\pp$, within evaluation noise). In a D4-only control arm, ablating D4 yields small but non-zero drops (63.6\% $\to$ 62.8\% or 62.4\%, i.e., 0.8--1.2$\pp$).

\paragraph{The D4 Paradox.} D4 improves training dynamics ($+3.4\pp$ over D0 at 5 epochs, and d4a4 outperforms A4-only by consistent margins) but shows weak or near-zero inference-time effect once training converges in top dual models. We hypothesize D4 acts primarily as a \emph{training regularizer}: by providing the MIDI encoder with explicit interval context early in training, it accelerates the discovery of useful representations that the encoder eventually internalizes.

\subsection{Transposition Invariance (Test 04)}
\label{sec:transposition}

A musically meaningful test of cross-modal representations is their robustness to pitch transposition: shifting all MIDI pitches by $\pm k$ semitones should preserve the musical content (modulo range effects) while the audio remains unchanged. We evaluate S at transpositions $k \in \{-6, -3, -1, 0, +1, +3, +6\}$ semitones (\cref{tab:transposition}, \cref{fig:transposition}).

\input{figures/transposition}

\begin{table}[t]
  \centering
  \caption{Transposition invariance (Test~04) for the seed-42 checkpoint. S (\%) at various semitone shifts. Retention = S$_{|\pm3|}$ / S$_0$. Multi-seed training replication is summarized in \cref{tab:multiseed}.}
  \label{tab:transposition}
  \begin{tabular}{lcccccccc}
    \toprule
    Arm & $-6$ & $-3$ & $-1$ & $0$ & $+1$ & $+3$ & $+6$ & Ret.\ $|\pm 3|$ \\
    \midrule
    D0     & 13.8 & 26.6 & 65.6 & 73.4 & 64.0 & 27.4 & 13.4 & 37\% \\
    d4a4   & 24.2 & 41.4 & 75.2 & 83.8 & 75.6 & 44.6 & 25.6 & 51\% \\
    \textbf{a4r}    & \textbf{27.0} & \textbf{46.2} & \textbf{76.6} & 82.0 & \textbf{76.8} & \textbf{51.0} & \textbf{27.6} & \textbf{59\%} \\
    d4-a4r & 27.0 & 45.0 & 73.2 & 79.8 & 75.2 & 49.2 & 27.2 & 59\% \\
    \bottomrule
  \end{tabular}
\end{table}

\paragraph{Key finding.} The reverse cross-attention models (a4r, d4-a4r) show the highest transposition invariance: at $\pm 3$ semitones, a4r retains 59\% of its baseline S compared to 37\% for D0---an advantage of $+23.6\pp$ in absolute S. This suggests that reverse cross-attention, by forcing the model to attend to encoder features \emph{from the perspective of spectral dynamics}, encourages learning of interval-relative representations that generalize across pitch shifts.

\subsection{Cross-Encoder Alignment --- CKA Analysis (Test 06)}
\label{sec:cka}

To understand \emph{how} descriptors improve performance, we measure the representational similarity between audio and MIDI transformer layers using Centered Kernel Alignment~\citep{kornblith2019cka}:
\begin{equation}
  \CKA(X, Y) = \frac{\|Y^\top X\|_F^2}{\|X^\top X\|_F \cdot \|Y^\top Y\|_F},
  \label{eq:cka}
\end{equation}
where $X \in \R^{n \times p}$ and $Y \in \R^{n \times q}$ are centered activation matrices for $n$ samples across $p$ and $q$ neurons respectively. CKA = 1 indicates identical representations (up to rotation and scaling); CKA = 0 indicates orthogonal representations.

We compute CKA for all 16 audio$\times$MIDI layer pairs (4 audio layers $\times$ 4 MIDI layers) using 500 matched pairs (\cref{tab:cka}, \cref{fig:cka_heatmaps}).

\input{figures/cka_heatmaps}

\begin{table}[t]
  \centering
  \caption{Cross-encoder CKA alignment (Test~06). Mean CKA across all 16 audio$\times$MIDI layer pairs. Higher = more aligned internal representations between encoders.}
  \label{tab:cka}
  \begin{tabular}{lccc}
    \toprule
    Arm & CKA mean & $\Delta$ vs D0 & RSA mean \\
    \midrule
    D0      & 0.435 & ---   & 0.446 \\
    d4a4    & 0.659 & $+51\%$ & 0.646 \\
    a4r     & 0.766 & $+76\%$ & 0.721 \\
    \textbf{d4-a4r}  & \textbf{0.794} & $\mathbf{+82\%}$ & \textbf{0.761} \\
    \bottomrule
  \end{tabular}
\end{table}

\paragraph{Descriptors double cross-encoder alignment.} The mean CKA increases monotonically with descriptor complexity: D0 (0.435) $\to$ d4a4 (0.659) $\to$ a4r (0.766) $\to$ d4-a4r (0.794). Descriptors cause the audio and MIDI transformers to develop \emph{similar internal representations}---they ``speak the same language.''

\paragraph{Alignment $\neq$ retrieval.} Critically, d4-a4r has the highest CKA (0.794) but \emph{not} the highest S (79.8\% vs.\ 83.8\% for d4a4). This dissociation suggests that while representational alignment is necessary for good cross-modal retrieval, \emph{excessive} alignment may reduce the capacity for modality-specific information that aids discrimination. The relationship between CKA and S appears to be non-monotonic.

\paragraph{Layer-depth analysis.} CKA increases near-monotonically with layer depth in all models (e.g., D0: 0.305 at layer~0$\times$0 to 0.722 at layer~3$\times$3; d4-a4r: 0.686 to 0.873), consistent with the expectation that deeper layers develop more abstract, modality-invariant representations.

\subsection{Spectral Band Sensitivity (Test 08)}
\label{sec:sensitivity}

We probe which A4 octave bands carry the most discriminative information by applying perturbations ($\epsilon = 0.1$) to individual bands and measuring the mean $L_2$ change in the output embedding (\cref{tab:sensitivity}, \cref{fig:sensitivity}).

\input{figures/sensitivity}

\begin{table}[t]
  \centering
  \caption{Perturbation sensitivity by A4 octave band (Test~08). Values are mean $\|\Delta z\|_2$ when perturbing each band by $\epsilon = 0.1$. Higher = more sensitive. Bands 4--7 (750+ Hz) dominate.}
  \label{tab:sensitivity}
  \begin{tabular}{clcccc}
    \toprule
    Band & Hz range & d4a4 & a4r & d4-a4r & Zone \\
    \midrule
    0 & 47\,--\,94       & 0.073 & 0.238 & 0.303 & Low \\
    1 & 94\,--\,188      & 0.224 & 0.335 & 0.313 & Low \\
    2 & 188\,--\,375     & 0.375 & 0.381 & 0.514 & Mid \\
    3 & 375\,--\,750     & 0.546 & 0.423 & 0.526 & Mid \\
    \midrule
    4 & 750\,--\,1500    & \textbf{0.664} & 0.478 & \textbf{0.773} & High \\
    5 & 1500\,--\,3000   & \textbf{0.662} & 0.476 & 0.599 & High \\
    6 & 3000\,--\,6000   & 0.264 & \textbf{0.875} & \textbf{1.092} & High \\
    7 & 6000\,--\,12000  & 0.209 & \textbf{0.933} & 0.529 & High \\
    \bottomrule
  \end{tabular}
\end{table}

\paragraph{High-frequency dominance.} In all models, bands 4--7 (750--12000\,Hz) carry 2--10$\times$ more discriminative power than bands 0--3. This is musically meaningful: the upper harmonic series ($>750$\,Hz) carries the most information about harmonic relationships and timbral identity, while low-frequency fundamentals are more uniform across pieces.

\paragraph{Mechanism-dependent spectral focus.} Each injection mechanism ``listens'' to different spectral regions: d4a4 (concatenation) is most sensitive to bands 4--5 (750--3000\,Hz), while a4r (reverse cross-attention) is most sensitive to bands 6--7 (3000--12000\,Hz). The d4-a4r combination peaks at band~6 (3000--6000\,Hz, sensitivity 1.092).

\paragraph{Non-linear encoding.} Despite high perturbation sensitivity, the Pearson correlation between A4 band values and output embedding dimensions is uniformly low ($|r| < 0.05$), indicating that the descriptor information is encoded through \emph{highly non-linear transformations} rather than being linearly preserved.

\subsection{Input Invariance (Test 09)}
\label{sec:invariance}

We assess robustness to four input perturbations: temporal misalignment, MIDI velocity scaling, octave transposition, and additive audio noise (\cref{tab:invariance,fig:invariance_noise}).

\begin{table}[t]
  \centering
  \caption{Perturbation robustness (Test~09). Each cell shows worst-case $S$ (\%) across perturbation directions at the given level. Temporal shift values are in samples (24\,kHz sample rate).}
  \label{tab:invariance}
  \begin{tabular}{llcccc}
    \toprule
    Perturbation & Level & D0 & d4a4 & a4r & d4-a4r \\
    \midrule
    Temporal shift & $\pm$8000\,samp ($\pm$0.33\,s) & 68.2 & 76.6 & 75.0 & 76.2 \\
    Velocity scale & 0.5$\times$ & 5.2 & 8.8 & 6.8 & 6.4 \\
    Velocity scale & 1.5$\times$ & 18.4 & 12.8 & 12.4 & 11.6 \\
    Octave transp. & $-$12\,st & 12.0 & 16.0 & 17.4 & 17.0 \\
    Octave transp. & $+$12\,st & 10.0 & 13.8 & 16.0 & 15.8 \\
    \bottomrule
  \end{tabular}
\end{table}

\paragraph{Temporal robustness.} At $\pm$8000 samples ($\pm$0.33\,s at 24\,kHz), worst-case drops range from $-3.6\pp$ (d4-a4r: $79.8 \to 76.2$) to $-7.2\pp$ (d4a4: $83.8 \to 76.6$). D0 drops $-5.2\pp$ ($73.4 \to 68.2$). All models retain $>90\%$ of baseline $S$, which is expected given that a 0.33\,s shift is ${\sim}8\%$ of the 4-second segment duration.

\paragraph{Velocity fragility.} All models collapse at 0.5$\times$ ($S = 5$--$9\%$) and 1.5$\times$ ($S = 12$--$18\%$). Even 0.8$\times$ causes large drops. The degradation is asymmetric: $S$ remains at 44--55\% absolute at 1.2$\times$ vs.\ only 32--47\% at 0.8$\times$. Note that velocity scaling modifies only MIDI velocity values, not note durations or onsets. The MIDI encoder learns a specific velocity$\to$embedding mapping during training; scaling distorts this mapping because the model has never seen velocity values outside the natural training distribution. The asymmetry may reflect two effects: at 0.5$\times$, many notes are pushed toward velocity $\approx 0$, collapsing dynamic range; at 1.5$\times$, values are clamped to 127, saturating the upper range. Velocity augmentation during training is a natural candidate for improving robustness.

\paragraph{Octave catastrophe.} All models collapse at $\pm$12 semitones ($S = 10$--$17\%$). The a4r and d4-a4r models retain slightly higher $S$ (16--17\%) than D0 and d4a4 (10--16\%), consistent with the transposition advantage seen in Test~04 at smaller intervals. Combined with Test~04 (\cref{sec:transposition}), this confirms that all models learn pitch-absolute representations; the partial invariance at $\pm$1--3 semitones does not extend to octave-scale shifts.

\paragraph{Audio noise: non-monotonic ranking.} D0 is most robust at moderate noise (20\,dB: $-0.4\pp$ vs.\ $-29\pp$ for d4a4), because it has no A4 descriptor pathway that can be corrupted. However, at 5\,dB the ranking reverses: d4-a4r (33.0\%) $>$ a4r (31.8\%) $>$ d4a4 (25.0\%) $>$ D0 (17.8\%). At moderate noise, descriptor corruption dominates (A4 is computed from noisy audio). At extreme noise, the MIDI encoder---which is noise-invariant by construction---provides a floor that is higher in descriptor-augmented models due to their stronger cross-modal alignment (CKA, Test~06). D0's MIDI encoder lacks this alignment and collapses when the audio side is destroyed. The mechanism-dependent spectral sensitivity (Test~08, \cref{sec:sensitivity}) may explain why a4r/d4-a4r are more sensitive at moderate noise---they rely more heavily on high-frequency bands (3000--12000\,Hz), which are first corrupted by additive Gaussian noise.

\input{figures/invariance}

\subsection{Linear Probing (Test 03)}
\label{sec:probing}

We evaluate the linear decodability of domain-specific features from frozen embeddings using 6 diagnostic probes (\cref{tab:probes}).

\begin{table}[t]
  \centering
  \caption{Linear probe $R^2$ from frozen 256-dim embeddings (Test~03). Cross-modal probes decode features of one modality from the other modality's embedding.}
  \label{tab:probes}
  \begin{tabular}{llcccc}
    \toprule
    Type & Probe & D0 & d4a4 & a4r & d4-a4r \\
    \midrule
    \multirow{4}{*}{Cross}
    & Audio $\to$ pitch histogram  & 0.181 & 0.174 & 0.167 & 0.186 \\
    & Audio $\to$ interval hist.   & 0.094 & 0.112 & 0.095 & 0.115 \\
    & \textbf{MIDI $\to$ chroma}   & \textbf{0.330} & 0.245 & 0.255 & 0.251 \\
    & MIDI $\to$ centroid          & 0.616 & 0.637 & \textbf{0.662} & 0.652 \\
    \midrule
    \multirow{2}{*}{Self}
    & Audio $\to$ chroma           & 0.310 & 0.235 & 0.249 & 0.231 \\
    & MIDI $\to$ pitch histogram   & 0.239 & 0.236 & 0.233 & 0.233 \\
    \bottomrule
  \end{tabular}
\end{table}

\paragraph{Counterintuitive result.} D0 (no descriptors) achieves the highest $R^2$ on the most informative cross-modal probe (MIDI $\to$ chroma: 0.330 vs.\ $\sim$0.25 for descriptor models). This suggests that descriptor-augmented models encode cross-modal information in a way that is \emph{less linearly decodable} but more effective for retrieval. The discriminative advantage of descriptors lives in distance geometry (cosine similarity structure) rather than in linearly accessible features.

\subsection{Embedding Space Visualization (Test 10)}
\label{sec:visualization}

We visualize the 256-dimensional embedding spaces using t-SNE~\citep{vandermaaten2008tsne} and UMAP~\citep{mcinnes2018umap} with 2000 samples per model. Audio and MIDI embeddings are plotted together, color-coded by modality, with lines connecting matched pairs. Additionally, we compute the distribution of cosine similarities between matched audio--MIDI pairs (\cref{fig:tsne_appendix,fig:umap_appendix,fig:cosine_align_appendix}).

The visualizations confirm quantitative findings: descriptor-augmented models show tighter clustering of matched pairs and higher mean cosine alignment (d4a4: 0.72 mean cosine vs.\ D0: 0.58). A compact summary dashboard of the seed-42 validation results is shown in \cref{fig:dashboard}; full t-SNE/UMAP details and cosine-alignment distribution are provided in \cref{app:visualizations}.

\input{figures/dashboard}

\subsection{Multi-Seed Training Replication}
\label{sec:multiseed}

The headline performance claims in this paper are anchored in a five-seed training replication over the four canonical arms. This summary should be read separately from the seed-42 diagnostic tables above: it reports training variability across independent reruns rather than within-checkpoint structure.

\begin{table}[t]
  \centering
  \caption{Multi-seed training replication across the four canonical arms. Means, standard deviations, and ranges are computed over five independent training seeds.}
  \label{tab:multiseed}
  \begin{tabular}{lcccc}
    \toprule
    Arm & Mean S (\%) & SD (pp) & Range & $\Delta$ vs D0 (pp) \\
    \midrule
    \textbf{d4a4}  & \textbf{84.0} & \textbf{2.7} & \textbf{81.4--87.6} & \textbf{+8.8} \\
    d4-a4r         & 81.2          & 2.5          & 78.4--83.4          & +6.0 \\
    a4r            & 80.7          & 1.9          & 79.4--84.0          & +5.5 \\
    D0             & 75.2          & 2.3          & 71.8--77.4          & --- \\
    \bottomrule
  \end{tabular}
\end{table}

The replicated ordering remains stable across reruns. For the flagship comparison, d4a4 exceeds D0 by $+8.8\pp$ on mean S, with a large effect size (Cohen~$d = 3.51$; Welch~$t = 5.55$; $p < 0.001$). The observed five-seed ranges do not overlap: the worst guided seed (81.4) remains 4.0 percentage points above the best unguided seed (77.4).

\section{Discussion}
\label{sec:discussion}

\subsection{The A4 Mechanism: Spectral Dynamics as Cross-Modal Bridge}

The most striking finding of this study is the complete causal dependence on A4---a relatively simple 8-dimensional descriptor capturing temporal changes in octave-band energy. The model does not merely \emph{use} A4 as supplementary information; it has reorganized its entire representational strategy around it. When A4 is removed, performance collapses to below random chance ($S < 8\%$), far worse than the 73.4\% achieved by D0 which never had access to A4.

This suggests that A4 provides a \emph{bridge representation}: a feature that is simultaneously computable from audio (via STFT) and predictable from MIDI (via expected harmonic content). Temporal dynamics of spectral energy---how energy in each octave band changes from frame to frame---capture the rhythmic and harmonic contour of a performance in a way that transcends the audio/symbolic divide.

\subsection{The D4 Paradox: Training Regularization Without Inference Causality}

D4 consistently improves screening performance ($+3.4\pp$ at 5 epochs) and contributes to d4a4 achieving the all-time record, yet in top dual models its ablation effect at inference is near-zero ($-0.4$ to $+0.6\pp$). In the D4-only control arm, the effect is small but non-zero (0.8--1.2$\pp$). We propose three complementary explanations:

\begin{enumerate}[nosep]
  \item \textbf{Curriculum effect}: D4 provides explicit interval context that helps the MIDI encoder discover useful features faster, but the encoder eventually learns equivalent representations from raw note sequences.
  \item \textbf{Gradient diversity}: D4 features create additional gradient pathways during backpropagation, improving optimization landscape navigation without remaining necessary in the forward pass.
  \item \textbf{Implicit regularization}: The projection from 516 to 512 dimensions (\cref{eq:concat}) acts as a bottleneck that encourages the encoder to compress descriptor information into its internal representations.
\end{enumerate}

\subsection{Reverse Cross-Attention: Informational Bottleneck as Feature}

In the canonical seed-42 comparison, the reverse cross-attention mechanism (a4r) achieves $S = 82.0\%$---competitive with the $83.8\%$ record of d4a4---while being $163\times$ cheaper in attention operations. Beyond computational efficiency, a4r shows the best transposition invariance (+23.6$\pp$ advantage over D0 at $\pm 3$ semitones). We attribute this to the \emph{informational bottleneck} created by using only 188 descriptor tokens as queries: by forcing all encoder information to flow through spectral-dynamics-shaped queries, the model is encouraged to learn representations structured around how energy changes across frequency bands---a property that is inherently invariant to absolute pitch.

\subsection{Alignment and Retrieval Are Not Monotonic}

The CKA analysis reveals a counterintuitive dissociation: d4-a4r has the highest cross-encoder alignment (CKA = 0.794) but not the highest retrieval performance ($S = 79.8\%$ vs.\ $83.8\%$ for d4a4 with CKA = 0.659). This suggests a non-monotonic relationship between representational alignment and task performance. We hypothesize that moderate alignment preserves sufficient modality-specific information for discrimination, while excessive alignment homogenizes representations at the cost of discriminative detail.

\subsection{Noise Robustness and Descriptor Vulnerability}

Descriptor-augmented models show larger relative $S$ drops than D0 at moderate noise levels ($\geq 20$\,dB SNR), but at 5\,dB the ranking reverses and descriptor models retain higher absolute $S$ (\cref{fig:invariance_noise}). The noise robustness profile is thus non-monotonic: A4 acts as both a strength and a vulnerability---it provides a powerful cross-modal bridge but creates a new attack surface through the audio-computed descriptor pathway. At moderate noise, A4 corruption dominates; at extreme noise, the superior cross-modal alignment of descriptor models (CKA 0.66--0.79 vs.\ 0.44 for D0, \cref{sec:cka}) provides residual matching ability through the noise-invariant MIDI pathway. Applications in noisy environments should consider the full robustness profile alongside peak performance.

\subsection{Limitations}

\begin{enumerate}[nosep]
  \item \textbf{Single dataset}: All experiments use MAESTRO v3.0.0, which contains exclusively classical piano. Generalization to other instruments, genres, or polyphonic settings is unknown.
  \item \textbf{Headline replication is now closed}: The flagship d4a4 arm has been replicated across five independent training seeds, yielding $84.0\% \pm 2.7\pp$. The observed five-seed ranges do not overlap with D0, and the effect size against the baseline remains large (Cohen~$d = 3.51$, $p < 0.001$).
  \item \textbf{Parameter-matched controls are now closed}: Test~02 isolates descriptor content from parameter-count inflation within the same architectural family. The real arm reaches 83.0\%, while the parameter-matched zero, random, and shuffled controls fall to 75.0\%, 73.6\%, and 73.6\%, respectively.
  \item \textbf{Audio encoder partially frozen}: The MERTEncoderLite architecture is partially initialized from pre-trained weights, potentially confounding the analysis of descriptor effects with pre-training biases.
  \item \textbf{No input augmentation}: All models are trained on unperturbed data. Velocity scaling and octave transposition cause catastrophic performance drops ($S < 18\%$), suggesting that data augmentation during training could substantially improve robustness (see Test~09, \cref{sec:invariance}).
\end{enumerate}

\subsection{Future Directions}

\begin{enumerate}[nosep]
  \item \textbf{Other instruments and genres}: Testing on datasets with broader timbral diversity (e.g., MusicNet~\citep{thickstun2017musicnet}).
  \item \textbf{Descriptor discovery}: Using the A4 success as a template for systematic descriptor search, potentially via learned descriptors initialized from hand-crafted ones.
  \item \textbf{Scaling}: Exploring whether the descriptor injection paradigm scales to larger encoders and datasets.
  \item \textbf{Robustness through augmentation}: Incorporating velocity scaling, octave transposition, and audio corruption during training to test whether the current fragilities can be reduced without erasing the descriptor-guided advantage.
\end{enumerate}

\section{Conclusion}
\label{sec:conclusion}

We have presented a systematic exploration of descriptor injection for cross-modal audio--MIDI learning, spanning 24 experimental configurations across 3 phases of increasing rigor. Our findings challenge the assumption that cross-modal improvement requires complex architectural innovations: a simple 8-dimensional spectral dynamics descriptor (A4), injected via concatenation, yields a replicated mean advantage of $+8.8$ percentage points over a strong VICReg baseline.

The scientific validation reveals that this improvement is not merely additive feature concatenation. Causal ablation shows the model has become \emph{entirely dependent} on A4, reorganizing its representations around spectral dynamics. CKA analysis reveals that the mechanism is \emph{representational convergence}: descriptors cause the audio and MIDI encoders to develop similar internal representations, effectively teaching both modalities a shared language grounded in spectral dynamics.

The reverse cross-attention mechanism offers both practical ($163\times$ cheaper) and scientific (enhanced transposition invariance) advantages, demonstrating that structured information bottlenecks can improve generalization. The D4 paradox---clear training benefit with weak inference-time causality in this model family---suggests that domain-specific descriptors may play a role as training regularizers even when their forward-pass contribution is limited at deployment.

These results position descriptor injection as a lightweight, effective paradigm for cross-modal learning in the music domain, and suggest that the principle may generalize to other cross-modal settings where domain knowledge can be encoded as structured features. This work is situated within a broader theoretical program---Harmonic Information Theory---which examines the hypothesis that frequency ratios may function as informational carriers across physical modalities. A companion volume develops the full theoretical framework~\citep{fernandezmendez2026hit}.

\paragraph{AI assistance disclosure.}
This research was conducted with the assistance of AI language models, including Claude (Anthropic), Codex, and other models, which served as coding assistants, documentation aids, experimental design advisors, and analytical tools throughout the project. All scientific decisions, experimental designs, and interpretations were made by the human author. The AI tools were used as assistants in multiple capacities: code implementation, data analysis, literature search, and manuscript preparation.

\begin{ack}
The author thanks Asociaci\'{o}n Civil AlterMundi for institutional support. This work used computational resources from UNC Superc\'omputo (CCAD) -- Universidad Nacional de C\'ordoba (\url{https://supercomputo.unc.edu.ar}), which are part of SNCAD, Rep\'ublica Argentina. Additional experiments were conducted on a local RTX~3090 workstation. The MAESTRO dataset~\citep{hawthorne2019maestro} was made available by the Magenta team at Google.
\end{ack}

\bibliographystyle{plainnat}
\bibliography{references}

\newpage

\input{appendix}

\end{document}

%% file: figures/architecture.tex
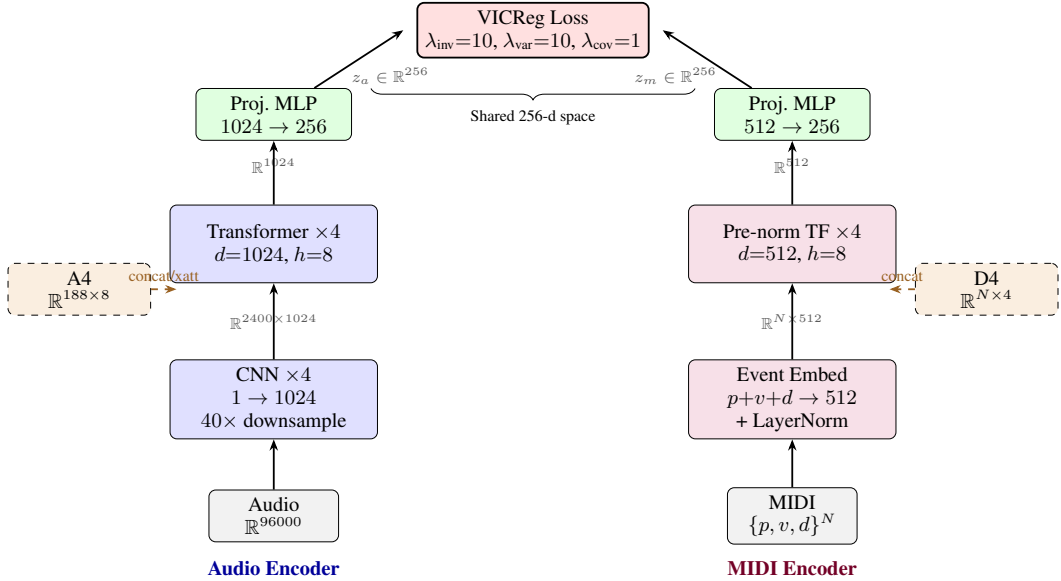
\begin{figure}[t]
  \centering
  \resizebox{\linewidth}{!}{%
  \begin{tikzpicture}[
    box/.style={draw, rounded corners=3pt, minimum height=0.8cm, minimum width=2.2cm, align=center, font=\small},
    encoder/.style={box, fill=blue!12, minimum width=3.2cm},
    proj/.style={box, fill=green!12, minimum width=2.4cm},
    desc/.style={box, fill=DAfour!15, minimum width=2.2cm, dashed},
    loss/.style={box, fill=red!12, minimum width=2.8cm, thick},
    arr/.style={-{Stealth[length=5pt]}, thick},
    darr/.style={-{Stealth[length=5pt]}, thick, DAfour!70!black, dashed},
    label/.style={font=\scriptsize\itshape, text=gray!70!black},
    >=Stealth
  ]
  \node[box, fill=gray!10, minimum width=2cm] (audio_in) at (-4, 0) {Audio\\$\R^{96000}$};
  \node[encoder, minimum height=1.2cm] (cnn) at (-4, 1.8) {CNN $\times 4$\\$1 \to 1024$\\$40\times$ downsample};
  \node[label] at (-4, 3.0) {$\R^{2400 \times 1024}$};
  \node[desc] (a4) at (-7, 3.5) {A4\\$\R^{188 \times 8}$};
  \node[encoder, minimum height=1.2cm] (audio_tf) at (-4, 4.2) {Transformer $\times 4$\\$d{=}1024$, $h{=}8$};
  \node[label] at (-4, 5.4) {$\R^{1024}$};
  \node[proj] (audio_proj) at (-4, 6.2) {Proj.\ MLP\\$1024 \to 256$};
  \node[box, fill=gray!10, minimum width=2cm] (midi_in) at (4, 0) {MIDI\\$\{p,v,d\}^N$};
  \node[encoder, minimum height=1.2cm, fill=purple!12] (embed) at (4, 1.8) {Event Embed\\$p{+}v{+}d \to 512$\\+ LayerNorm};
  \node[label] at (4, 3.0) {$\R^{N \times 512}$};
  \node[desc] (d4) at (7, 3.5) {D4\\$\R^{N \times 4}$};
  \node[encoder, minimum height=1.2cm, fill=purple!12] (midi_tf) at (4, 4.2) {Pre-norm TF $\times 4$\\$d{=}512$, $h{=}8$};
  \node[label] at (4, 5.4) {$\R^{512}$};
  \node[proj] (midi_proj) at (4, 6.2) {Proj.\ MLP\\$512 \to 256$};
  \node[loss] (vicreg) at (0, 7.5) {VICReg Loss\\$\lambda_{\text{inv}}{=}10$, $\lambda_{\text{var}}{=}10$, $\lambda_{\text{cov}}{=}1$};
  \node[label] at (-2.2, 6.8) {$z_a \in \R^{256}$};
  \node[label] at (2.2, 6.8) {$z_m \in \R^{256}$};
  \draw[arr] (audio_in) -- (cnn);
  \draw[arr] (cnn) -- (audio_tf);
  \draw[arr] (audio_tf) -- (audio_proj);
  \draw[arr] (audio_proj) -- (-2, 7.5);
  \draw[arr] (midi_in) -- (embed);
  \draw[arr] (embed) -- (midi_tf);
  \draw[arr] (midi_tf) -- (midi_proj);
  \draw[arr] (midi_proj) -- (2, 7.5);
  \draw[darr] (a4) -- (-5.5, 3.5) node[midway, above, font=\scriptsize, DAfour!70!black] {concat/xatt};
  \draw[darr] (d4) -- (5.5, 3.5) node[midway, above, font=\scriptsize, DAfour!70!black] {concat};
  \node[font=\footnotesize\bfseries, blue!60!black] at (-4, -0.8) {Audio Encoder};
  \node[font=\footnotesize\bfseries, purple!60!black] at (4, -0.8) {MIDI Encoder};
  \draw[decorate, decoration={brace, amplitude=5pt, mirror}] (-2.5, 6.6) -- (2.5, 6.6) node[midway, below=6pt, font=\scriptsize] {Shared 256-d space};

  \end{tikzpicture}%
  }
  \caption{Model architecture for the canonical audio--MIDI baseline. The audio branch uses a 4-layer CNN plus a 4-layer Transformer; the MIDI branch uses event embeddings plus a 4-layer pre-norm Transformer. Both project to a shared 256-D space and train with VICReg. Dashed arrows mark optional descriptor injection points (A4 on audio, D4 on MIDI).}
  \label{fig:architecture}
\end{figure}

%% file: figures/screening.tex
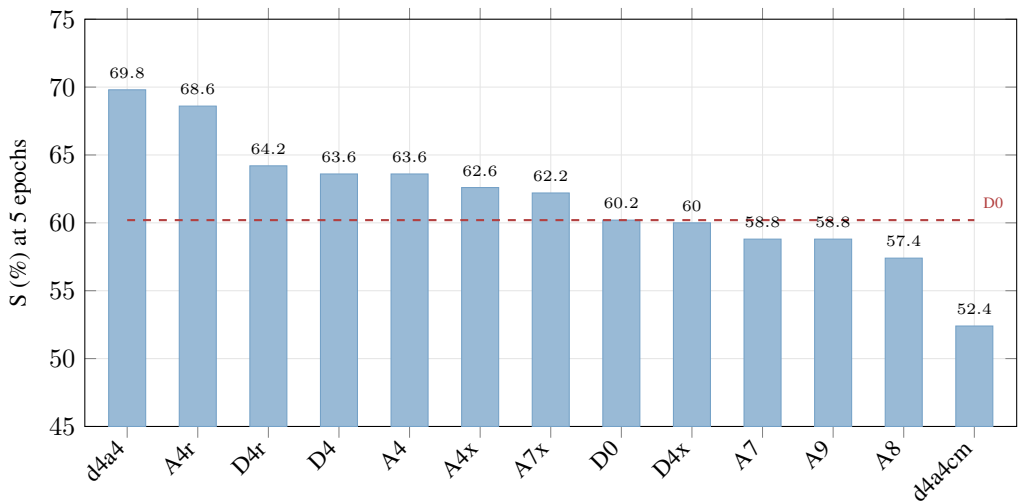
\begin{figure}[t]
  \centering
  \begin{tikzpicture}
    \begin{axis}[
      ybar,
      width=\textwidth,
      height=7cm,
      bar width=14pt,
      ylabel={S (\%) at 5 epochs},
      symbolic x coords={d4a4, A4r, D4r, D4, A4, A4x, A7x, D0, D4x, A7, A9, A8, d4a4cm},
      xtick=data,
      x tick label style={rotate=45, anchor=east, font=\small},
      ymin=45,
      ymax=75,
      ytick={45,50,55,60,65,70,75},
      nodes near coords,
      nodes near coords style={font=\tiny, above},
      every node near coord/.append style={yshift=1pt},
      enlarge x limits=0.05,
      grid=major,
      grid style={gray!20},
      ylabel style={font=\small},
    ]

    \addplot[
      fill=Dfourfill, draw=Dfourdraw,
    ] coordinates {
      (d4a4, 69.8)
      (A4r, 68.6)
      (D4r, 64.2)
      (D4, 63.6)
      (A4, 63.6)
      (A4x, 62.6)
      (A7x, 62.2)
      (D0, 60.2)
      (D4x, 60.0)
      (A7, 58.8)
      (A9, 58.8)
      (A8, 57.4)
      (d4a4cm, 52.4)
    };
    \draw[dashed, negcol, thick] (axis cs:d4a4, 60.2) -- (axis cs:d4a4cm, 60.2);
    \node[font=\tiny, negcol, anchor=west] at (axis cs:d4a4cm, 61.5) {D0};

    \end{axis}
  \end{tikzpicture}
  \caption{Gate~4.3 screening results: 13 descriptor$\times$mechanism combinations ranked by S at 5 training epochs. Dashed line marks D0 baseline ($S{=}60.2\%$). The top 7 arms all inject descriptors into their own modality's encoder. Cross-modal injection (d4a4cm) is worst.}
  \label{fig:screening}
\end{figure}

%% file: figures/ablation.tex
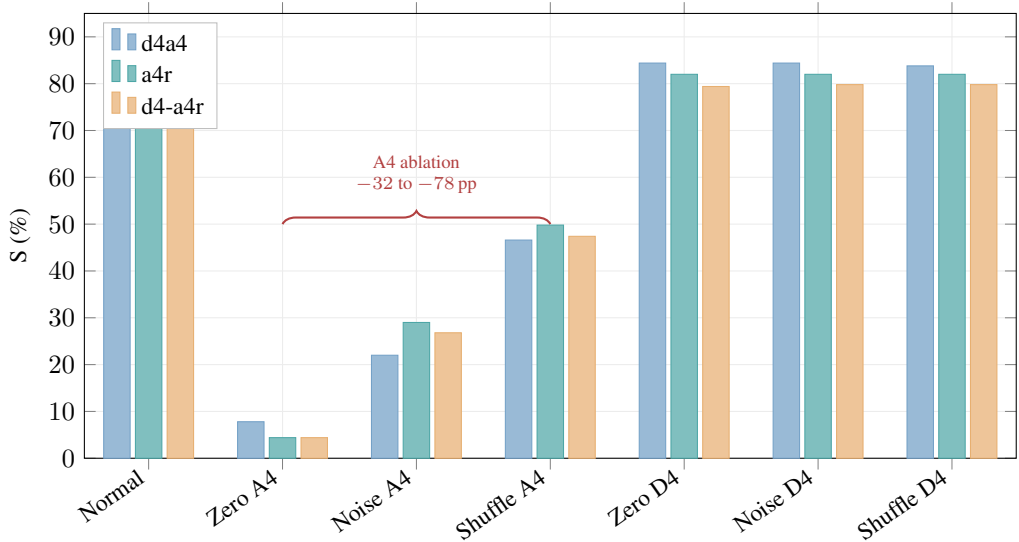
\begin{figure}[t]
  \centering
  \begin{tikzpicture}
    \begin{axis}[
      ybar,
      width=\textwidth,
      height=7.5cm,
      bar width=10pt,
      ylabel={S (\%)},
      symbolic x coords={Normal, Zero A4, Noise A4, Shuffle A4, Zero D4, Noise D4, Shuffle D4},
      xtick=data,
      x tick label style={rotate=35, anchor=east, font=\small},
      ymin=0,
      ymax=95,
      ytick={0,10,20,30,40,50,60,70,80,90},
      legend style={
        at={(0.02,0.98)},
        anchor=north west,
        font=\small,
        cells={anchor=west},
        draw=gray!50,
      },
      enlarge x limits=0.08,
      grid=major,
      grid style={gray!15},
      ylabel style={font=\small},
    ]
    \addplot[fill=Dfourfill, draw=Dfourdraw] coordinates {
      (Normal, 83.8)
      (Zero A4, 7.8)
      (Noise A4, 22.0)
      (Shuffle A4, 46.6)
      (Zero D4, 84.4)
      (Noise D4, 84.4)
      (Shuffle D4, 83.8)
    };
    \addplot[fill=Afourfill, draw=Afourdraw] coordinates {
      (Normal, 82.0)
      (Zero A4, 4.4)
      (Noise A4, 29.0)
      (Shuffle A4, 49.8)
      (Zero D4, 82.0)
      (Noise D4, 82.0)
      (Shuffle D4, 82.0)
    };
    \addplot[fill=DAfourfill, draw=DAfourdraw] coordinates {
      (Normal, 79.8)
      (Zero A4, 4.4)
      (Noise A4, 26.8)
      (Shuffle A4, 47.4)
      (Zero D4, 79.4)
      (Noise D4, 79.8)
      (Shuffle D4, 79.8)
    };

    \legend{d4a4, a4r, d4-a4r}
    \draw[decorate, decoration={brace, amplitude=5pt}, negcol, thick]
      (axis cs:Zero A4, 50) -- (axis cs:Shuffle A4, 50)
      node[midway, above=8pt, font=\scriptsize, negcol, align=center] {A4 ablation\\$-32$ to $-78\pp$};
    \draw[decorate, decoration={brace, amplitude=5pt, mirror}, poscol, thick]
      (axis cs:Zero D4, -2) -- (axis cs:Shuffle D4, -2)
      node[midway, below=8pt, font=\scriptsize, poscol, align=center] {D4 ablation\\${\sim}0\pp$};

    \end{axis}
  \end{tikzpicture}
  \caption{Causal ablation results (Test~01) for the canonical seed-42 checkpoint. A4 ablation (left group) collapses S by 32--78$\pp$ across all models. D4 ablation (right group) has near-zero effect in dual models ($-0.4$ to $+0.6\pp$). Note: a4r has no D4 descriptor, so its D4 bars equal its Normal bar for visual alignment.}
  \label{fig:ablation}
\end{figure}

%% file: figures/transposition.tex
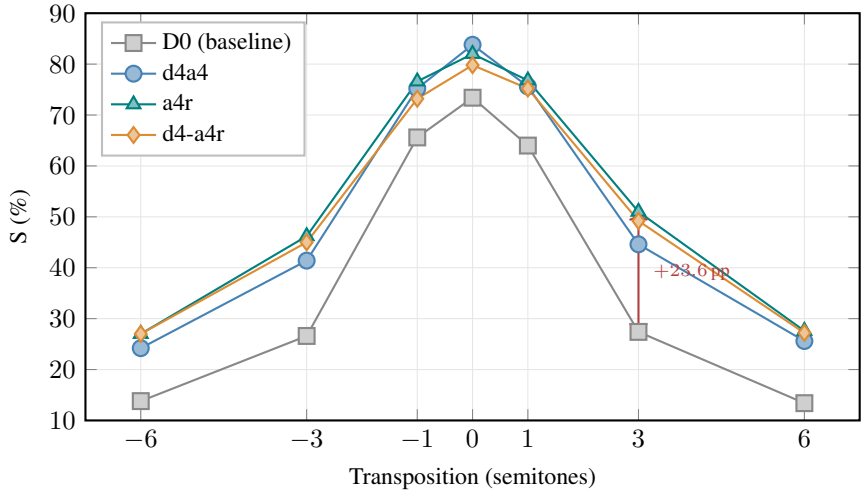
\begin{figure}[t]
  \centering
  \begin{tikzpicture}
    \begin{axis}[
      width=0.85\textwidth,
      height=7cm,
      xlabel={Transposition (semitones)},
      ylabel={S (\%)},
      xmin=-7, xmax=7,
      ymin=10, ymax=90,
      xtick={-6,-3,-1,0,1,3,6},
      ytick={10,20,30,40,50,60,70,80,90},
      grid=major,
      grid style={gray!20},
      legend style={
        at={(0.02,0.98)},
        anchor=north west,
        font=\small,
        cells={anchor=west},
        draw=gray!50,
      },
      xlabel style={font=\small},
      ylabel style={font=\small},
      mark size=3pt,
      thick,
    ]
    \addplot[color=Dzero, mark=square*, mark options={fill=Dzerofill}] coordinates {
      (-6, 13.8) (-3, 26.6) (-1, 65.6) (0, 73.4) (1, 64.0) (3, 27.4) (6, 13.4)
    };
    \addplot[color=Dfour, mark=*, mark options={fill=Dfourfill}] coordinates {
      (-6, 24.2) (-3, 41.4) (-1, 75.2) (0, 83.8) (1, 75.6) (3, 44.6) (6, 25.6)
    };
    \addplot[color=Afour, mark=triangle*, mark options={fill=Afourfill}] coordinates {
      (-6, 27.0) (-3, 46.2) (-1, 76.6) (0, 82.0) (1, 76.8) (3, 51.0) (6, 27.6)
    };
    \addplot[color=DAfour, mark=diamond*, mark options={fill=DAfourfill}] coordinates {
      (-6, 27.0) (-3, 45.0) (-1, 73.2) (0, 79.8) (1, 75.2) (3, 49.2) (6, 27.2)
    };

    \legend{D0 (baseline), d4a4, a4r, d4-a4r}
    \draw[<->, thick, negcol] (axis cs:3, 27.4) -- (axis cs:3, 51.0)
      node[midway, right=2pt, font=\scriptsize, negcol] {$+23.6\pp$};

    \end{axis}
  \end{tikzpicture}
  \caption{Transposition invariance (Test~04) for the canonical seed-42 checkpoint. All models degrade with pitch shift, but descriptor-augmented models retain more performance. At $\pm 3$ semitones, a4r shows $+23.6\pp$ advantage over D0. The reverse cross-attention models (a4r, d4-a4r) are the most invariant.}
  \label{fig:transposition}
\end{figure}

%% file: figures/cka_heatmaps.tex
\newcommand{\ckacell}[5]{%
  \pgfmathsetmacro{\pctA}{max(0, min(100, #3*100))}%
  \fill[yellow!\pctA!violet!80!blue] (#1*0.88+#4, #2*0.88+#5) rectangle ++(0.88,0.88);
  \node[minimum size=8.8mm, inner sep=0pt, font=\scriptsize, text=black]
    at (#1*0.88+#4+0.44, #2*0.88+#5+0.44)
    {\pgfmathprintnumber[fixed, fixed zerofill, precision=2]{#3}};
}
\newcommand{\ckalabels}[2]{%
  \foreach \i/\lab in {0/A0, 1/A1, 2/A2, 3/A3, 4/M0, 5/M1, 6/M2, 7/M3} {
    \node[font=\scriptsize, anchor=east] at (#1-0.08, \i*0.88+#2+0.44) {\lab};
    \node[font=\scriptsize, anchor=north] at (\i*0.88+#1+0.44, #2-0.08) {\lab};
  }
}

\begin{figure}[t]
  \centering
  \resizebox{\textwidth}{!}{%
  \begin{tikzpicture}
  \def\xA{0}\def\yA{8.5}
  \node[font=\small\bfseries] at (\xA+3.52, \yA+7.6) {D0~~(S=73.4\%)};
  \ckacell{0}{7}{1.00}{\xA}{\yA} \ckacell{1}{7}{0.86}{\xA}{\yA} \ckacell{2}{7}{0.58}{\xA}{\yA} \ckacell{3}{7}{0.45}{\xA}{\yA} \ckacell{4}{7}{0.31}{\xA}{\yA} \ckacell{5}{7}{0.21}{\xA}{\yA} \ckacell{6}{7}{0.13}{\xA}{\yA} \ckacell{7}{7}{0.13}{\xA}{\yA}
  \ckacell{0}{6}{0.86}{\xA}{\yA} \ckacell{1}{6}{1.00}{\xA}{\yA} \ckacell{2}{6}{0.78}{\xA}{\yA} \ckacell{3}{6}{0.56}{\xA}{\yA} \ckacell{4}{6}{0.40}{\xA}{\yA} \ckacell{5}{6}{0.32}{\xA}{\yA} \ckacell{6}{6}{0.21}{\xA}{\yA} \ckacell{7}{6}{0.20}{\xA}{\yA}
  \ckacell{0}{5}{0.58}{\xA}{\yA} \ckacell{1}{5}{0.78}{\xA}{\yA} \ckacell{2}{5}{1.00}{\xA}{\yA} \ckacell{3}{5}{0.88}{\xA}{\yA} \ckacell{4}{5}{0.54}{\xA}{\yA} \ckacell{5}{5}{0.63}{\xA}{\yA} \ckacell{6}{5}{0.60}{\xA}{\yA} \ckacell{7}{5}{0.57}{\xA}{\yA}
  \ckacell{0}{4}{0.45}{\xA}{\yA} \ckacell{1}{4}{0.56}{\xA}{\yA} \ckacell{2}{4}{0.88}{\xA}{\yA} \ckacell{3}{4}{1.00}{\xA}{\yA} \ckacell{4}{4}{0.54}{\xA}{\yA} \ckacell{5}{4}{0.72}{\xA}{\yA} \ckacell{6}{4}{0.74}{\xA}{\yA} \ckacell{7}{4}{0.72}{\xA}{\yA}
  \ckacell{0}{3}{0.31}{\xA}{\yA} \ckacell{1}{3}{0.40}{\xA}{\yA} \ckacell{2}{3}{0.54}{\xA}{\yA} \ckacell{3}{3}{0.54}{\xA}{\yA} \ckacell{4}{3}{1.00}{\xA}{\yA} \ckacell{5}{3}{0.77}{\xA}{\yA} \ckacell{6}{3}{0.53}{\xA}{\yA} \ckacell{7}{3}{0.47}{\xA}{\yA}
  \ckacell{0}{2}{0.21}{\xA}{\yA} \ckacell{1}{2}{0.32}{\xA}{\yA} \ckacell{2}{2}{0.63}{\xA}{\yA} \ckacell{3}{2}{0.72}{\xA}{\yA} \ckacell{4}{2}{0.77}{\xA}{\yA} \ckacell{5}{2}{1.00}{\xA}{\yA} \ckacell{6}{2}{0.88}{\xA}{\yA} \ckacell{7}{2}{0.83}{\xA}{\yA}
  \ckacell{0}{1}{0.13}{\xA}{\yA} \ckacell{1}{1}{0.21}{\xA}{\yA} \ckacell{2}{1}{0.60}{\xA}{\yA} \ckacell{3}{1}{0.74}{\xA}{\yA} \ckacell{4}{1}{0.53}{\xA}{\yA} \ckacell{5}{1}{0.88}{\xA}{\yA} \ckacell{6}{1}{1.00}{\xA}{\yA} \ckacell{7}{1}{0.98}{\xA}{\yA}
  \ckacell{0}{0}{0.13}{\xA}{\yA} \ckacell{1}{0}{0.20}{\xA}{\yA} \ckacell{2}{0}{0.57}{\xA}{\yA} \ckacell{3}{0}{0.72}{\xA}{\yA} \ckacell{4}{0}{0.47}{\xA}{\yA} \ckacell{5}{0}{0.83}{\xA}{\yA} \ckacell{6}{0}{0.98}{\xA}{\yA} \ckacell{7}{0}{1.00}{\xA}{\yA}
  \draw[poscol, dashed, line width=0.8pt] (4*0.88+\xA, 4*0.88+\yA) rectangle (8*0.88+\xA, 8*0.88+\yA);
  \draw[black!30, line width=0.3pt] (4*0.88+\xA, 0+\yA) -- (4*0.88+\xA, 8*0.88+\yA);
  \draw[black!30, line width=0.3pt] (0+\xA, 4*0.88+\yA) -- (8*0.88+\xA, 4*0.88+\yA);
  \ckalabels{\xA}{\yA}
  \def\xB{8.5}\def\yB{8.5}
  \node[font=\small\bfseries] at (\xB+3.52, \yB+7.6) {d4a4~~(S=83.8\%)};
  \ckacell{0}{7}{1.00}{\xB}{\yB} \ckacell{1}{7}{0.95}{\xB}{\yB} \ckacell{2}{7}{0.73}{\xB}{\yB} \ckacell{3}{7}{0.60}{\xB}{\yB} \ckacell{4}{7}{0.47}{\xB}{\yB} \ckacell{5}{7}{0.50}{\xB}{\yB} \ckacell{6}{7}{0.46}{\xB}{\yB} \ckacell{7}{7}{0.42}{\xB}{\yB}
  \ckacell{0}{6}{0.95}{\xB}{\yB} \ckacell{1}{6}{1.00}{\xB}{\yB} \ckacell{2}{6}{0.89}{\xB}{\yB} \ckacell{3}{6}{0.77}{\xB}{\yB} \ckacell{4}{6}{0.59}{\xB}{\yB} \ckacell{5}{6}{0.65}{\xB}{\yB} \ckacell{6}{6}{0.63}{\xB}{\yB} \ckacell{7}{6}{0.58}{\xB}{\yB}
  \ckacell{0}{5}{0.73}{\xB}{\yB} \ckacell{1}{5}{0.89}{\xB}{\yB} \ckacell{2}{5}{1.00}{\xB}{\yB} \ckacell{3}{5}{0.93}{\xB}{\yB} \ckacell{4}{5}{0.69}{\xB}{\yB} \ckacell{5}{5}{0.80}{\xB}{\yB} \ckacell{6}{5}{0.81}{\xB}{\yB} \ckacell{7}{5}{0.76}{\xB}{\yB}
  \ckacell{0}{4}{0.60}{\xB}{\yB} \ckacell{1}{4}{0.77}{\xB}{\yB} \ckacell{2}{4}{0.93}{\xB}{\yB} \ckacell{3}{4}{1.00}{\xB}{\yB} \ckacell{4}{4}{0.69}{\xB}{\yB} \ckacell{5}{4}{0.81}{\xB}{\yB} \ckacell{6}{4}{0.86}{\xB}{\yB} \ckacell{7}{4}{0.83}{\xB}{\yB}
  \ckacell{0}{3}{0.47}{\xB}{\yB} \ckacell{1}{3}{0.59}{\xB}{\yB} \ckacell{2}{3}{0.69}{\xB}{\yB} \ckacell{3}{3}{0.69}{\xB}{\yB} \ckacell{4}{3}{1.00}{\xB}{\yB} \ckacell{5}{3}{0.91}{\xB}{\yB} \ckacell{6}{3}{0.80}{\xB}{\yB} \ckacell{7}{3}{0.69}{\xB}{\yB}
  \ckacell{0}{2}{0.50}{\xB}{\yB} \ckacell{1}{2}{0.65}{\xB}{\yB} \ckacell{2}{2}{0.80}{\xB}{\yB} \ckacell{3}{2}{0.81}{\xB}{\yB} \ckacell{4}{2}{0.91}{\xB}{\yB} \ckacell{5}{2}{1.00}{\xB}{\yB} \ckacell{6}{2}{0.94}{\xB}{\yB} \ckacell{7}{2}{0.85}{\xB}{\yB}
  \ckacell{0}{1}{0.46}{\xB}{\yB} \ckacell{1}{1}{0.63}{\xB}{\yB} \ckacell{2}{1}{0.81}{\xB}{\yB} \ckacell{3}{1}{0.86}{\xB}{\yB} \ckacell{4}{1}{0.80}{\xB}{\yB} \ckacell{5}{1}{0.94}{\xB}{\yB} \ckacell{6}{1}{1.00}{\xB}{\yB} \ckacell{7}{1}{0.95}{\xB}{\yB}
  \ckacell{0}{0}{0.42}{\xB}{\yB} \ckacell{1}{0}{0.58}{\xB}{\yB} \ckacell{2}{0}{0.76}{\xB}{\yB} \ckacell{3}{0}{0.83}{\xB}{\yB} \ckacell{4}{0}{0.69}{\xB}{\yB} \ckacell{5}{0}{0.85}{\xB}{\yB} \ckacell{6}{0}{0.95}{\xB}{\yB} \ckacell{7}{0}{1.00}{\xB}{\yB}
  \draw[poscol, dashed, line width=0.8pt] (4*0.88+\xB, 4*0.88+\yB) rectangle (8*0.88+\xB, 8*0.88+\yB);
  \draw[black!30, line width=0.3pt] (4*0.88+\xB, 0+\yB) -- (4*0.88+\xB, 8*0.88+\yB);
  \draw[black!30, line width=0.3pt] (0+\xB, 4*0.88+\yB) -- (8*0.88+\xB, 4*0.88+\yB);
  \ckalabels{\xB}{\yB}
  \def\xC{0}\def\yC{0}
  \node[font=\small\bfseries] at (\xC+3.52, \yC+7.6) {a4r~~(S=82.0\%)};
  \ckacell{0}{7}{1.00}{\xC}{\yC} \ckacell{1}{7}{0.98}{\xC}{\yC} \ckacell{2}{7}{0.93}{\xC}{\yC} \ckacell{3}{7}{0.86}{\xC}{\yC} \ckacell{4}{7}{0.65}{\xC}{\yC} \ckacell{5}{7}{0.74}{\xC}{\yC} \ckacell{6}{7}{0.74}{\xC}{\yC} \ckacell{7}{7}{0.73}{\xC}{\yC}
  \ckacell{0}{6}{0.98}{\xC}{\yC} \ckacell{1}{6}{1.00}{\xC}{\yC} \ckacell{2}{6}{0.97}{\xC}{\yC} \ckacell{3}{6}{0.92}{\xC}{\yC} \ckacell{4}{6}{0.65}{\xC}{\yC} \ckacell{5}{6}{0.76}{\xC}{\yC} \ckacell{6}{6}{0.79}{\xC}{\yC} \ckacell{7}{6}{0.78}{\xC}{\yC}
  \ckacell{0}{5}{0.93}{\xC}{\yC} \ckacell{1}{5}{0.97}{\xC}{\yC} \ckacell{2}{5}{1.00}{\xC}{\yC} \ckacell{3}{5}{0.98}{\xC}{\yC} \ckacell{4}{5}{0.69}{\xC}{\yC} \ckacell{5}{5}{0.82}{\xC}{\yC} \ckacell{6}{5}{0.85}{\xC}{\yC} \ckacell{7}{5}{0.84}{\xC}{\yC}
  \ckacell{0}{4}{0.86}{\xC}{\yC} \ckacell{1}{4}{0.92}{\xC}{\yC} \ckacell{2}{4}{0.98}{\xC}{\yC} \ckacell{3}{4}{1.00}{\xC}{\yC} \ckacell{4}{4}{0.67}{\xC}{\yC} \ckacell{5}{4}{0.81}{\xC}{\yC} \ckacell{6}{4}{0.87}{\xC}{\yC} \ckacell{7}{4}{0.86}{\xC}{\yC}
  \ckacell{0}{3}{0.65}{\xC}{\yC} \ckacell{1}{3}{0.65}{\xC}{\yC} \ckacell{2}{3}{0.69}{\xC}{\yC} \ckacell{3}{3}{0.67}{\xC}{\yC} \ckacell{4}{3}{1.00}{\xC}{\yC} \ckacell{5}{3}{0.92}{\xC}{\yC} \ckacell{6}{3}{0.76}{\xC}{\yC} \ckacell{7}{3}{0.70}{\xC}{\yC}
  \ckacell{0}{2}{0.74}{\xC}{\yC} \ckacell{1}{2}{0.76}{\xC}{\yC} \ckacell{2}{2}{0.82}{\xC}{\yC} \ckacell{3}{2}{0.81}{\xC}{\yC} \ckacell{4}{2}{0.92}{\xC}{\yC} \ckacell{5}{2}{1.00}{\xC}{\yC} \ckacell{6}{2}{0.92}{\xC}{\yC} \ckacell{7}{2}{0.88}{\xC}{\yC}
  \ckacell{0}{1}{0.74}{\xC}{\yC} \ckacell{1}{1}{0.79}{\xC}{\yC} \ckacell{2}{1}{0.85}{\xC}{\yC} \ckacell{3}{1}{0.87}{\xC}{\yC} \ckacell{4}{1}{0.76}{\xC}{\yC} \ckacell{5}{1}{0.92}{\xC}{\yC} \ckacell{6}{1}{1.00}{\xC}{\yC} \ckacell{7}{1}{0.98}{\xC}{\yC}
  \ckacell{0}{0}{0.73}{\xC}{\yC} \ckacell{1}{0}{0.78}{\xC}{\yC} \ckacell{2}{0}{0.84}{\xC}{\yC} \ckacell{3}{0}{0.86}{\xC}{\yC} \ckacell{4}{0}{0.70}{\xC}{\yC} \ckacell{5}{0}{0.88}{\xC}{\yC} \ckacell{6}{0}{0.98}{\xC}{\yC} \ckacell{7}{0}{1.00}{\xC}{\yC}
  \draw[poscol, dashed, line width=0.8pt] (4*0.88+\xC, 4*0.88+\yC) rectangle (8*0.88+\xC, 8*0.88+\yC);
  \draw[black!30, line width=0.3pt] (4*0.88+\xC, 0+\yC) -- (4*0.88+\xC, 8*0.88+\yC);
  \draw[black!30, line width=0.3pt] (0+\xC, 4*0.88+\yC) -- (8*0.88+\xC, 4*0.88+\yC);
  \ckalabels{\xC}{\yC}
  \def\xD{8.5}\def\yD{0}
  \node[font=\small\bfseries] at (\xD+3.52, \yD+7.6) {d4-a4r~~(S=79.8\%)};
  \ckacell{0}{7}{1.00}{\xD}{\yD} \ckacell{1}{7}{0.98}{\xD}{\yD} \ckacell{2}{7}{0.91}{\xD}{\yD} \ckacell{3}{7}{0.85}{\xD}{\yD} \ckacell{4}{7}{0.69}{\xD}{\yD} \ckacell{5}{7}{0.74}{\xD}{\yD} \ckacell{6}{7}{0.75}{\xD}{\yD} \ckacell{7}{7}{0.74}{\xD}{\yD}
  \ckacell{0}{6}{0.98}{\xD}{\yD} \ckacell{1}{6}{1.00}{\xD}{\yD} \ckacell{2}{6}{0.97}{\xD}{\yD} \ckacell{3}{6}{0.92}{\xD}{\yD} \ckacell{4}{6}{0.72}{\xD}{\yD} \ckacell{5}{6}{0.80}{\xD}{\yD} \ckacell{6}{6}{0.81}{\xD}{\yD} \ckacell{7}{6}{0.80}{\xD}{\yD}
  \ckacell{0}{5}{0.91}{\xD}{\yD} \ckacell{1}{5}{0.97}{\xD}{\yD} \ckacell{2}{5}{1.00}{\xD}{\yD} \ckacell{3}{5}{0.98}{\xD}{\yD} \ckacell{4}{5}{0.76}{\xD}{\yD} \ckacell{5}{5}{0.85}{\xD}{\yD} \ckacell{6}{5}{0.87}{\xD}{\yD} \ckacell{7}{5}{0.85}{\xD}{\yD}
  \ckacell{0}{4}{0.85}{\xD}{\yD} \ckacell{1}{4}{0.92}{\xD}{\yD} \ckacell{2}{4}{0.98}{\xD}{\yD} \ckacell{3}{4}{1.00}{\xD}{\yD} \ckacell{4}{4}{0.74}{\xD}{\yD} \ckacell{5}{4}{0.84}{\xD}{\yD} \ckacell{6}{4}{0.88}{\xD}{\yD} \ckacell{7}{4}{0.87}{\xD}{\yD}
  \ckacell{0}{3}{0.69}{\xD}{\yD} \ckacell{1}{3}{0.72}{\xD}{\yD} \ckacell{2}{3}{0.76}{\xD}{\yD} \ckacell{3}{3}{0.74}{\xD}{\yD} \ckacell{4}{3}{1.00}{\xD}{\yD} \ckacell{5}{3}{0.94}{\xD}{\yD} \ckacell{6}{3}{0.84}{\xD}{\yD} \ckacell{7}{3}{0.76}{\xD}{\yD}
  \ckacell{0}{2}{0.74}{\xD}{\yD} \ckacell{1}{2}{0.80}{\xD}{\yD} \ckacell{2}{2}{0.85}{\xD}{\yD} \ckacell{3}{2}{0.84}{\xD}{\yD} \ckacell{4}{2}{0.94}{\xD}{\yD} \ckacell{5}{2}{1.00}{\xD}{\yD} \ckacell{6}{2}{0.96}{\xD}{\yD} \ckacell{7}{2}{0.90}{\xD}{\yD}
  \ckacell{0}{1}{0.75}{\xD}{\yD} \ckacell{1}{1}{0.81}{\xD}{\yD} \ckacell{2}{1}{0.87}{\xD}{\yD} \ckacell{3}{1}{0.88}{\xD}{\yD} \ckacell{4}{1}{0.84}{\xD}{\yD} \ckacell{5}{1}{0.96}{\xD}{\yD} \ckacell{6}{1}{1.00}{\xD}{\yD} \ckacell{7}{1}{0.98}{\xD}{\yD}
  \ckacell{0}{0}{0.74}{\xD}{\yD} \ckacell{1}{0}{0.80}{\xD}{\yD} \ckacell{2}{0}{0.85}{\xD}{\yD} \ckacell{3}{0}{0.87}{\xD}{\yD} \ckacell{4}{0}{0.76}{\xD}{\yD} \ckacell{5}{0}{0.90}{\xD}{\yD} \ckacell{6}{0}{0.98}{\xD}{\yD} \ckacell{7}{0}{1.00}{\xD}{\yD}
  \draw[poscol, dashed, line width=0.8pt] (4*0.88+\xD, 4*0.88+\yD) rectangle (8*0.88+\xD, 8*0.88+\yD);
  \draw[black!30, line width=0.3pt] (4*0.88+\xD, 0+\yD) -- (4*0.88+\xD, 8*0.88+\yD);
  \draw[black!30, line width=0.3pt] (0+\xD, 4*0.88+\yD) -- (8*0.88+\xD, 4*0.88+\yD);
  \ckalabels{\xD}{\yD}
  \begin{scope}[shift={(16.2, 0.5)}]
    \foreach \y/\val in {0/0.0, 1/0.1, 2/0.2, 3/0.3, 4/0.4, 5/0.5, 6/0.6, 7/0.7, 8/0.8, 9/0.9, 10/1.0} {
      \pgfmathsetmacro{\pctA}{min(100, \val*100)}%
      \fill[yellow!\pctA!violet!80!blue] (0, \y*0.7) rectangle (0.35, \y*0.7+0.7);
    }
    \draw (0,0) rectangle (0.35, 7.7);
    \foreach \y/\lab in {0/0.0, 2.5/0.25, 5/0.50, 7.5/0.75, 10/1.0} {
      \pgfmathsetmacro{\ypos}{\y*0.7}
      \node[font=\scriptsize, anchor=west] at (0.45, \ypos) {\lab};
    }
    \node[font=\scriptsize, rotate=90, anchor=south] at (-0.3, 3.85) {CKA (Linear)};
  \end{scope}
  \node[font=\scriptsize, text=gray!70!black, anchor=north] at (8.75, -0.5)
    {Green dashed boxes = cross-encoder block (Audio layers $\times$ MIDI layers).  Higher values = more similar representations.};

  \end{tikzpicture}%
  }
  \caption{CKA representational alignment across all 8$\times$8 layer pairs (Test~06) for the canonical seed-42 checkpoint. Rows/columns: audio layers (A0--A3) and MIDI layers (M0--M3). Green dashed boxes highlight the cross-encoder block. Within-encoder alignment is uniformly high ($>0.85$); cross-encoder alignment reveals the descriptor effect: D0 mean $= 0.435$; d4a4 $= 0.659$ ($+51\%$); a4r $= 0.766$ ($+76\%$); d4-a4r $= 0.794$ ($+82\%$). Alignment increases monotonically with layer depth in all models.}
  \label{fig:cka_heatmaps}
\end{figure}
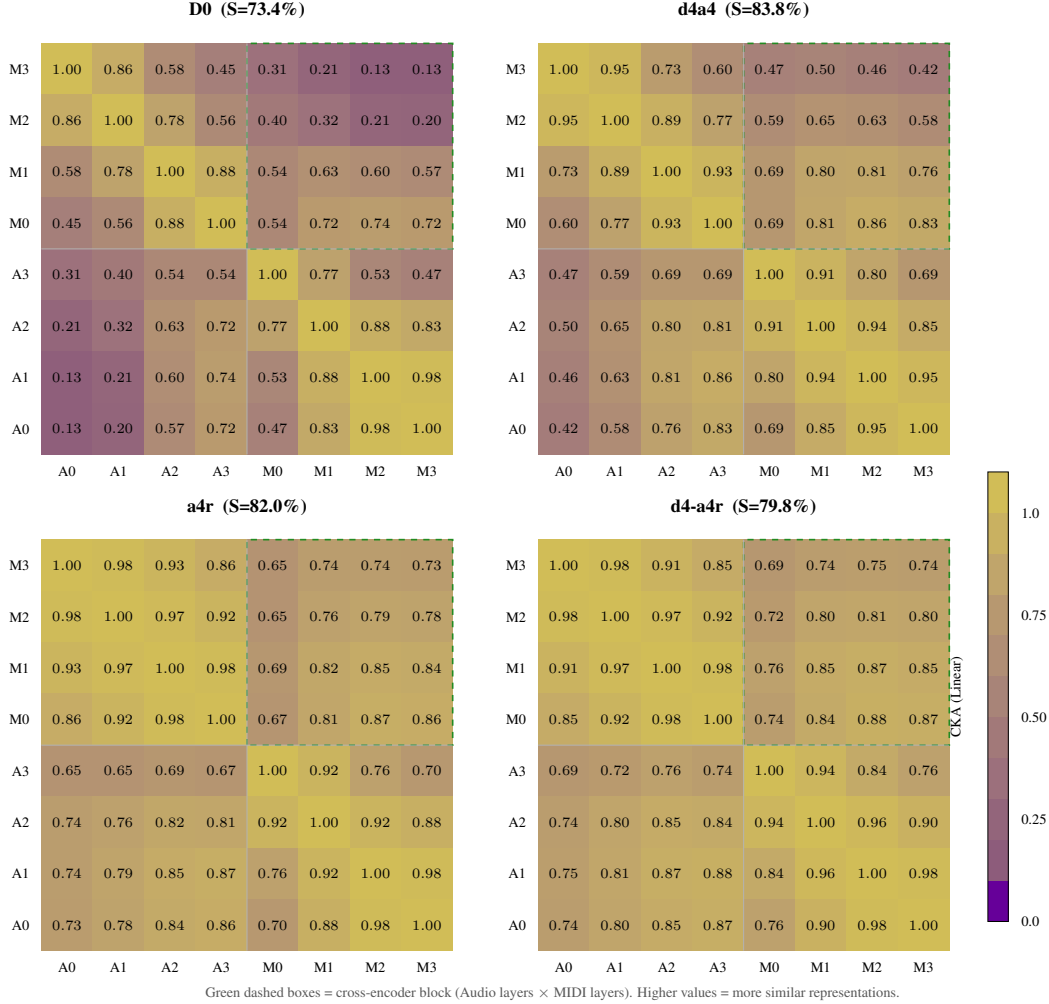

%% file: figures/sensitivity.tex
\begin{figure}[t]
  \centering
  \begin{tikzpicture}
    \begin{axis}[
      ybar,
      width=\textwidth,
      height=7cm,
      bar width=8pt,
      ylabel={Mean $\|\Delta z\|_2$ (sensitivity)},
      symbolic x coords={B0, B1, B2, B3, B4, B5, B6, B7},
      xtick=data,
      xticklabels={
        {47--94},
        {94--188},
        {188--375},
        {375--750},
        {750--1.5k},
        {1.5--3k},
        {3--6k},
        {6--12k}
      },
      x tick label style={rotate=45, anchor=east, font=\small},
      ymin=0,
      ymax=1.2,
      ytick={0, 0.2, 0.4, 0.6, 0.8, 1.0, 1.2},
      legend style={
        at={(0.02,0.98)},
        anchor=north west,
        font=\small,
        cells={anchor=west},
        draw=gray!50,
      },
      enlarge x limits=0.08,
      grid=major,
      grid style={gray!15},
      ylabel style={font=\small},
    ]
    \addplot[fill=Dfourfill, draw=Dfourdraw] coordinates {
      (B0, 0.073) (B1, 0.224) (B2, 0.375) (B3, 0.546)
      (B4, 0.664) (B5, 0.662) (B6, 0.264) (B7, 0.209)
    };
    \addplot[fill=Afourfill, draw=Afourdraw] coordinates {
      (B0, 0.238) (B1, 0.335) (B2, 0.381) (B3, 0.423)
      (B4, 0.478) (B5, 0.476) (B6, 0.875) (B7, 0.933)
    };
    \addplot[fill=DAfourfill, draw=DAfourdraw] coordinates {
      (B0, 0.303) (B1, 0.313) (B2, 0.514) (B3, 0.526)
      (B4, 0.773) (B5, 0.599) (B6, 1.092) (B7, 0.529)
    };

    \legend{d4a4, a4r, d4-a4r}
    \draw[dashed, gray!50, thick] (axis cs:B3, 0) -- (axis cs:B3, 1.15);
    \node[font=\scriptsize, gray!60] at (axis cs:B1, 1.12) {Low freq};
    \node[font=\scriptsize, gray!60] at (axis cs:B5, 1.12) {High freq};

    \end{axis}
  \end{tikzpicture}
  \caption{Perturbation sensitivity by A4 octave band (Test~08) for the canonical seed-42 checkpoint. X-axis labels show full band ranges in kHz/Hz. Each group shows the mean $L_2$ change in output embedding when perturbing one band by $\epsilon{=}0.1$. High-frequency bands (750+ Hz) carry 2--10$\times$ more sensitivity. Each injection mechanism attends to different spectral regions: d4a4 peaks at 750--3000\,Hz, a4r at 3--12\,kHz, d4-a4r at 3--6\,kHz.}
  \label{fig:sensitivity}
\end{figure}
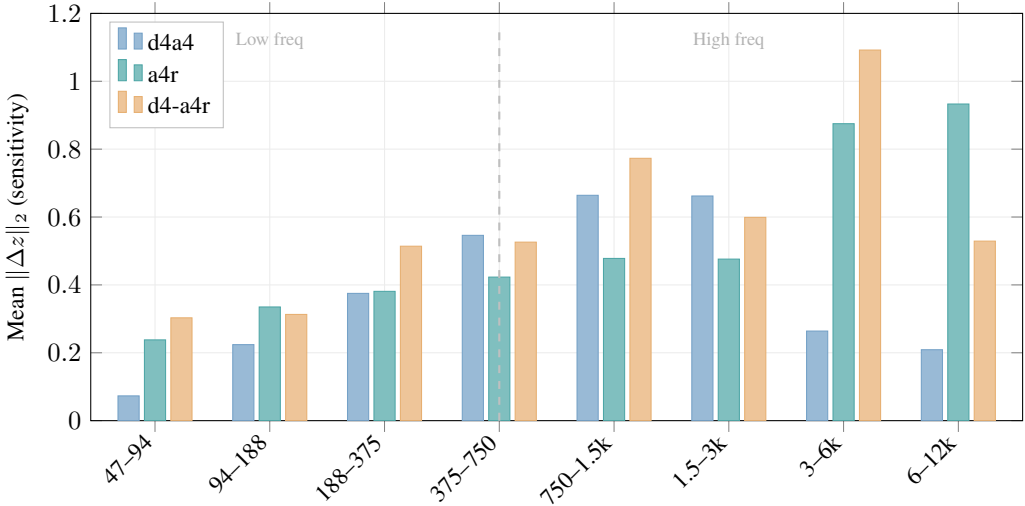

%% file: figures/invariance.tex
\begin{figure}[t]
  \centering
  \begin{tikzpicture}
    \begin{axis}[
      width=0.85\textwidth,
      height=7cm,
      xlabel={SNR (dB)},
      ylabel={S (\%)},
      xmin=-2, xmax=45,
      ymin=10, ymax=90,
      xtick={5,10,20,30,40},
      xticklabels={5,10,20,30,40},
      extra x ticks={44},
      extra x tick labels={clean},
      extra x tick style={grid=none},
      ytick={10,20,30,40,50,60,70,80,90},
      grid=major,
      grid style={gray!20},
      x dir=reverse,
      legend style={
        at={(0.02,0.02)},
        anchor=south west,
        font=\small,
        cells={anchor=west},
        draw=gray!50,
      },
      xlabel style={font=\small},
      ylabel style={font=\small},
      mark size=3pt,
      thick,
    ]
    \addplot[color=Dzero, mark=square*, mark options={fill=Dzerofill}] coordinates {
      (44, 73.4) (40, 73.4) (30, 73.4) (20, 73.0) (10, 46.8) (5, 17.8)
    };
    \addplot[color=Dfour, mark=*, mark options={fill=Dfourfill}] coordinates {
      (44, 83.8) (40, 79.8) (30, 67.0) (20, 54.8) (10, 52.2) (5, 25.0)
    };
    \addplot[color=Afour, mark=triangle*, mark options={fill=Afourfill}] coordinates {
      (44, 82.0) (40, 66.0) (30, 47.2) (20, 41.2) (10, 40.4) (5, 31.8)
    };
    \addplot[color=DAfour, mark=diamond*, mark options={fill=DAfourfill}] coordinates {
      (44, 79.8) (40, 67.6) (30, 50.2) (20, 40.4) (10, 41.8) (5, 33.0)
    };

    \legend{D0 (baseline), d4a4, a4r, d4-a4r}
    \draw[->, thick, poscol] (axis cs:7, 20) -- (axis cs:5.5, 31)
      node[pos=0, below, font=\scriptsize, poscol] {ranking reversal};

    \end{axis}
  \end{tikzpicture}
  \caption{Audio noise degradation (Test~09) for the canonical seed-42 checkpoint. At moderate noise ($\geq 20$\,dB), D0 is most robust because it has no corruptible A4 pathway. At 5\,dB the ranking reverses: descriptor models retain higher absolute $S$, likely due to their stronger cross-modal alignment providing a noise-invariant floor via the MIDI encoder.}
  \label{fig:invariance_noise}
\end{figure}
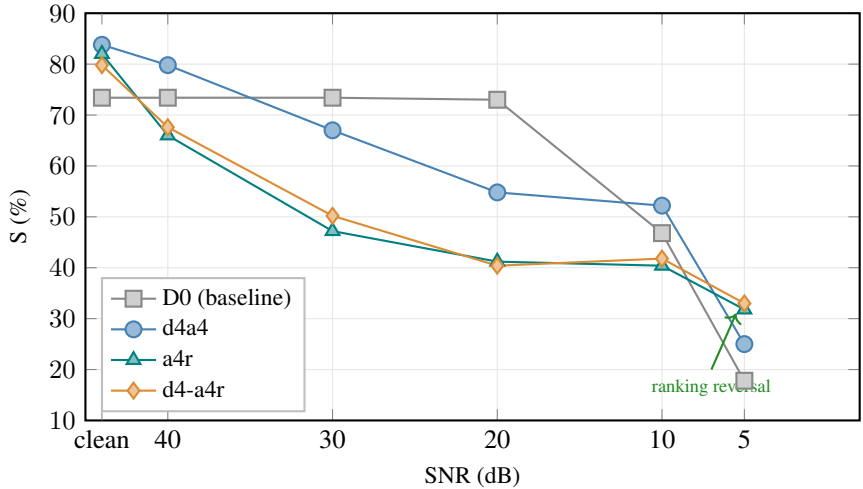

%% file: figures/dashboard.tex
\begin{figure}[t]
  \centering
  \resizebox{\linewidth}{!}{%
  \begin{tikzpicture}
    \begin{groupplot}[
      group style={
        group size=3 by 2,
        horizontal sep=1.6cm,
        vertical sep=2.0cm,
      },
      width=5.2cm,
      height=4.5cm,
      grid=major,
      grid style={gray!15},
      tick label style={font=\tiny},
      label style={font=\small},
      title style={font=\small\bfseries},
    ]
    \nextgroupplot[
      title={Retrieval (S)},
      ybar,
      bar width=12pt,
      symbolic x coords={D0, d4a4, a4r, d4-a4r},
      xtick=data,
      x tick label style={rotate=30, anchor=east},
      ymin=65, ymax=90,
      ylabel={S (\%)},
      ylabel style={font=\tiny},
      nodes near coords,
      nodes near coords style={font=\tiny},
    ]
    \addplot[fill=Dzerofill, draw=Dzerodraw] coordinates {(D0, 73.4)};
    \addplot[fill=Dfourfill, draw=Dfourdraw] coordinates {(d4a4, 83.8)};
    \addplot[fill=Afourfill, draw=Afourdraw] coordinates {(a4r, 82.0)};
    \addplot[fill=DAfourfill, draw=DAfourdraw] coordinates {(d4-a4r, 79.8)};
    \nextgroupplot[
      title={A4 Ablation (zero)},
      ybar,
      bar width=12pt,
      symbolic x coords={d4a4, a4r, d4-a4r},
      xtick=data,
      x tick label style={rotate=30, anchor=east},
      ymin=0, ymax=90,
      ylabel={S (\%)},
      ylabel style={font=\tiny},
    ]
    \addplot[fill=Dfourdesat, draw=Dfourdesatdraw] coordinates {(d4a4, 83.8)};
    \addplot[fill=Afourdesat, draw=Afourdesatdraw] coordinates {(a4r, 82.0)};
    \addplot[fill=DAfourdesat, draw=DAfourdesatdraw] coordinates {(d4-a4r, 79.8)};
    \addplot[fill=negcol!40, draw=negcol!60] coordinates {(d4a4, 7.8)};
    \addplot[fill=negcol!40, draw=negcol!60] coordinates {(a4r, 4.4)};
    \addplot[fill=negcol!40, draw=negcol!60] coordinates {(d4-a4r, 4.4)};
    \nextgroupplot[
      title={CKA Alignment},
      ybar,
      bar width=12pt,
      symbolic x coords={D0, d4a4, a4r, d4-a4r},
      xtick=data,
      x tick label style={rotate=30, anchor=east},
      ymin=0.3, ymax=0.9,
      ylabel={CKA mean},
      ylabel style={font=\tiny},
      nodes near coords,
      nodes near coords style={font=\tiny},
    ]
    \addplot[fill=Dzerofill, draw=Dzerodraw] coordinates {(D0, 0.435)};
    \addplot[fill=Dfourfill, draw=Dfourdraw] coordinates {(d4a4, 0.659)};
    \addplot[fill=Afourfill, draw=Afourdraw] coordinates {(a4r, 0.766)};
    \addplot[fill=DAfourfill, draw=DAfourdraw] coordinates {(d4-a4r, 0.794)};
    \nextgroupplot[
      title={Retention at $\pm 3$ st},
      ybar,
      bar width=12pt,
      symbolic x coords={D0, d4a4, a4r, d4-a4r},
      xtick=data,
      x tick label style={rotate=30, anchor=east},
      ymin=25, ymax=65,
      ylabel={Retention (\%)},
      ylabel style={font=\tiny},
      nodes near coords,
      nodes near coords style={font=\tiny},
    ]
    \addplot[fill=Dzerofill, draw=Dzerodraw] coordinates {(D0, 37)};
    \addplot[fill=Dfourfill, draw=Dfourdraw] coordinates {(d4a4, 51)};
    \addplot[fill=Afourfill, draw=Afourdraw] coordinates {(a4r, 59)};
    \addplot[fill=DAfourfill, draw=DAfourdraw] coordinates {(d4-a4r, 59)};
    \nextgroupplot[
      title={Peak Sensitivity},
      ybar,
      bar width=12pt,
      symbolic x coords={d4a4, a4r, d4-a4r},
      xtick=data,
      x tick label style={rotate=30, anchor=east},
      ymin=0, ymax=1.2,
      ylabel={Max $\|\Delta z\|_2$},
      ylabel style={font=\tiny},
      nodes near coords,
      nodes near coords style={font=\tiny},
    ]
    \addplot[fill=Dfourfill, draw=Dfourdraw] coordinates {(d4a4, 0.664)};
    \addplot[fill=Afourfill, draw=Afourdraw] coordinates {(a4r, 0.933)};
    \addplot[fill=DAfourfill, draw=DAfourdraw] coordinates {(d4-a4r, 1.092)};
    \nextgroupplot[
      title={Recall@1},
      ybar,
      bar width=10pt,
      symbolic x coords={D0, d4a4, a4r, d4-a4r},
      xtick=data,
      x tick label style={rotate=30, anchor=east},
      ymin=15, ymax=35,
      ylabel={R@1 (\%)},
      ylabel style={font=\tiny},
      nodes near coords,
      nodes near coords style={font=\tiny},
    ]
    \addplot[fill=Dfour!30, draw=Dfour!50] coordinates {(D0, 22.2) (d4a4, 28.0) (a4r, 27.6) (d4-a4r, 25.2)};
    \addplot[fill=Afour!30, draw=Afour!50] coordinates {(D0, 21.2) (d4a4, 27.6) (a4r, 25.8) (d4-a4r, 24.0)};

    \end{groupplot}
  \end{tikzpicture}%
  }
  \caption{Seed-42 summary dashboard: retrieval $S$, collapse under A4 ablation, CKA alignment, transposition retention at $\pm 3$ semitones, peak perturbation sensitivity, and bidirectional Recall@1.}
  \label{fig:dashboard}
\end{figure}
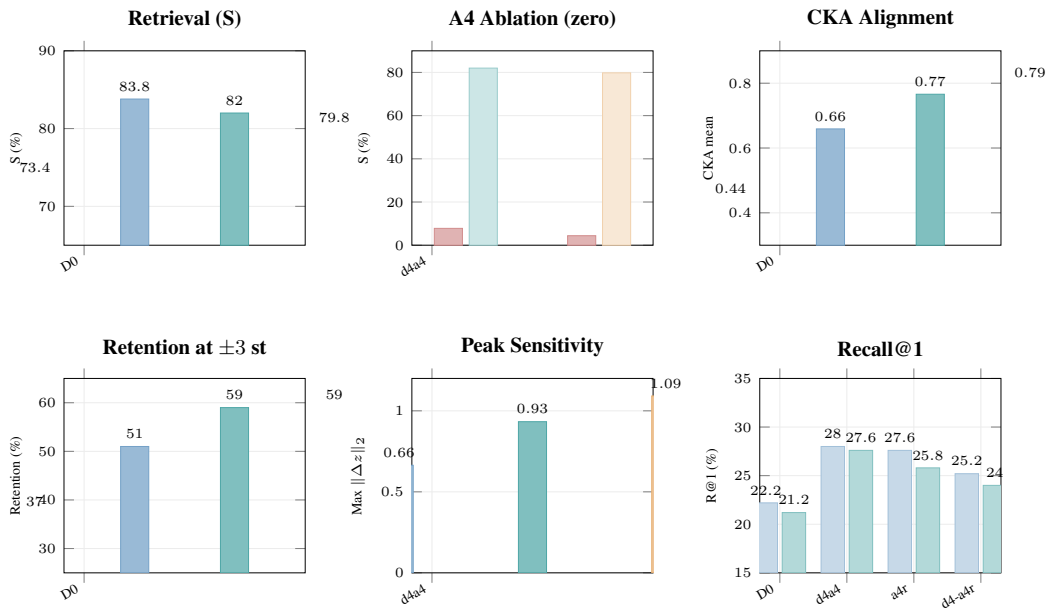

%% file: appendix.tex
\appendix

\section{Complete Descriptor Catalog}
\label{app:descriptors}

\cref{tab:descriptor_catalog} lists all 13 descriptor$\times$mechanism combinations explored during the screening phase. Below we provide full mathematical formulations for descriptors not covered in the main text.

\begin{table}[ht]
  \centering
  \caption{Complete descriptor catalog. Domain: A = audio, M = MIDI, D = dual. Mechanism: C = concatenation, X = standard cross-attention, R = reverse cross-attention, CM = cross-modal injection.}
  \label{tab:descriptor_catalog}
  \scriptsize
  \setlength{\tabcolsep}{4pt}
  \begin{tabular}{@{}lllcccp{4.2cm}@{}}
    \toprule
    Arm & Descriptor & Domain & Dims & Mechanism & S$_{5\text{ep}}$ (\%) & Description \\
    \midrule
    D0      & ---  & ---  & ---  & ---  & 60.2 & VICReg baseline (no descriptors) \\
    D4      & D4   & MIDI & 4    & C    & 63.6 & Local pitch intervals \\
    A4      & A4   & Audio & 8   & C    & 63.6 & Octave-band energy dynamics \\
    A7      & A7   & Audio & 12  & C    & 58.8 & Rational attractor (just intonation) \\
    A8      & A8   & Audio & 12  & C    & 57.4 & Onset-weighted chroma \\
    A9      & A9   & Audio & 12  & C    & 58.8 & IDF-weighted rational attractor \\
    D4x     & D4   & MIDI & 4    & X    & 60.0 & D4 via standard cross-attention \\
    A4x     & A4   & Audio & 8   & X    & 62.6 & A4 via standard cross-attention \\
    A7x     & A7   & Audio & 12  & X    & 62.2 & A7 via standard cross-attention \\
    D4r     & D4   & MIDI & 4    & R    & 64.2 & D4 via reverse cross-attention \\
    A4r     & A4   & Audio & 8   & R    & 68.6 & A4 via reverse cross-attention \\
    d4a4    & D4+A4 & Dual & 4+8 & C+C  & 69.8 & D4 concat (MIDI) + A4 concat (audio) \\
    d4a4cm  & D4+A4 & Dual & 4+8 & CM   & 52.4 & Cross-modal: A4$\to$MIDI, D4$\to$audio \\
    \bottomrule
  \end{tabular}
\end{table}

\subsection{A7: Rational Attractor Descriptor}
\label{app:a7}

The A7 descriptor projects spectral peak ratios onto 12 just-intonation attractors, capturing the harmonic structure of the signal in a musically meaningful basis.

\paragraph{Just-intonation attractors.} We define 12 target ratios corresponding to the just-intonation intervals within one octave (\cref{tab:attractors}).

\begin{table}[ht]
  \centering
  \caption{A7 just-intonation attractors. Log$_2$ values are octave-folded to $[0, 1)$.}
  \label{tab:attractors}
  \small
  \begin{tabular}{ccccl}
    \toprule
    Index & Ratio & log$_2$ & Cents & Musical interval \\
    \midrule
    0  & 1/1   & 0.000 & 0    & Unison / Octave \\
    1  & 16/15 & 0.093 & 112  & Minor second \\
    2  & 9/8   & 0.170 & 204  & Major second \\
    3  & 6/5   & 0.263 & 316  & Minor third \\
    4  & 5/4   & 0.322 & 386  & Major third \\
    5  & 4/3   & 0.415 & 498  & Perfect fourth \\
    6  & 7/5   & 0.485 & 583  & Tritone (septimal) \\
    7  & 3/2   & 0.585 & 702  & Perfect fifth \\
    8  & 8/5   & 0.678 & 814  & Minor sixth \\
    9  & 5/3   & 0.737 & 884  & Major sixth \\
    10 & 7/4   & 0.807 & 969  & Harmonic seventh \\
    11 & 15/8  & 0.907 & 1088 & Major seventh \\
    \bottomrule
  \end{tabular}
\end{table}

\paragraph{Algorithm.}
\begin{enumerate}[nosep]
  \item Compute STFT magnitude $|X[k,t]|$ with $n_{\text{FFT}} = 2048$, hop $= 512$.
  \item Extract top-$K = 8$ spectral peaks per frame by magnitude, masking peaks below 50\,Hz.
  \item Compute all $\binom{8}{2} = 28$ pairwise log-frequency ratios:
  \begin{equation}
    r_{ij}[t] = \log_2\!\left(\frac{f_j[t]}{f_i[t]}\right) \bmod 1.0, \quad j > i,
  \end{equation}
  folding into the octave range $[0, 1)$.
  \item Soft assignment to attractors via Gaussian kernel ($\sigma = 0.02$):
  \begin{equation}
    w_{ij,c}[t] = \sqrt{|X_{f_i}[t]| \cdot |X_{f_j}[t]|} \cdot \exp\!\left(-\frac{(r_{ij}[t] - \mu_c)^2}{2\sigma^2}\right),
  \end{equation}
  where $\mu_c$ is the log$_2$ value of attractor $c$.
  \item Accumulate over all pairs and normalize:
  \begin{equation}
    A7[t, c] = \frac{\sum_{i<j} w_{ij,c}[t]}{\sum_{c'}\sum_{i<j} w_{ij,c'}[t] + \epsilon}.
  \end{equation}
\end{enumerate}

Output: $A7 \in \R^{B \times T_{\text{stft}} \times 12}$, interpolated to $T = 2400$ for concatenation injection.

\subsection{A8: Onset-Weighted Chroma}
\label{app:a8}

The A8 descriptor computes a chroma representation weighted by onset strength, emphasizing transient events where note onsets provide the strongest pitch information.

\begin{enumerate}[nosep]
  \item Compute STFT magnitude $|X[k,t]|$.
  \item Map STFT bins to 12 pitch classes via log-frequency:
  \begin{equation}
    \text{pc}(k) = \left\lfloor 12 \cdot \log_2\!\left(\frac{k \cdot f_s / n_{\text{FFT}}}{f_{\text{ref}}}\right) \right\rfloor \bmod 12,
  \end{equation}
  where $f_{\text{ref}} = 32.7$\,Hz (C1).
  \item Accumulate energy per pitch class: $C[t, c] = \sum_{k:\,\text{pc}(k)=c} |X[k,t]|^2$.
  \item Compute spectral flux (onset strength):
  \begin{equation}
    F[t] = \sum_k \max\!\big(0,\; |X[k,t]| - |X[k,t{-}1]|\big).
  \end{equation}
  \item Normalize flux to $[0, 1]$ and gate chroma: $A8[t, c] = C[t, c] \cdot F[t] / \max(F)$.
  \item Normalize per frame: $A8[t, :] \leftarrow A8[t, :] / (\sum_c A8[t, c] + \epsilon)$.
\end{enumerate}

Output: $A8 \in \R^{B \times T_{\text{stft}} \times 12}$.

\subsection{A9: IDF-Weighted Rational Attractor}
\label{app:a9}

A9 extends A7 with inverse document frequency (IDF) weighting to suppress ubiquitous ratio activations:
\begin{enumerate}[nosep]
  \item Compute raw A7 activations $A7_{\text{raw}}[t, c]$ (before normalization).
  \item Compute document frequency: $\text{df}_c = |\{t : A7_{\text{raw}}[t, c] > \tau\}| / T$, with $\tau = 0.05$.
  \item Compute IDF: $\text{idf}_c = \text{clamp}\!\big(\log(1 / (\text{df}_c + 10^{-3})),\; 0,\; 5\big)$.
  \item Weight and normalize: $A9[t, c] = A7_{\text{raw}}[t, c] \cdot \text{idf}_c / \sum_{c'} A7_{\text{raw}}[t, c'] \cdot \text{idf}_{c'}$.
\end{enumerate}

The IDF weighting suppresses ratio activations that appear in $>50\%$ of frames (e.g., octave/unison), emphasizing rarer harmonic relationships.

\section{Architecture Details}
\label{app:architecture}

\subsection{MERTEncoderLite --- Full Specification}

\begin{table}[ht]
  \centering
  \caption{MERTEncoderLite CNN feature extractor.}
  \label{tab:cnn_spec}
  \begin{tabular}{cccccccc}
    \toprule
    Layer & In ch.\ & Out ch.\ & Kernel & Stride & Padding & Norm & Activation \\
    \midrule
    1 & 1    & 512  & 10 & 5 & 0 & GN(32) & GELU \\
    2 & 512  & 512  & 3  & 2 & 1 & GN(32) & GELU \\
    3 & 512  & 512  & 3  & 2 & 1 & GN(32) & GELU \\
    4 & 512  & 1024 & 3  & 2 & 1 & GN(32) & GELU \\
    \bottomrule
  \end{tabular}
\end{table}

Total downsampling: $5 \times 2 \times 2 \times 2 = 40\times$. Input $96{,}000 \to$ output $2{,}400$ frames.

\begin{table}[ht]
  \centering
  \caption{Audio Transformer encoder specification.}
  \label{tab:audio_transformer}
  \begin{tabular}{lc}
    \toprule
    Hyperparameter & Value \\
    \midrule
    Number of layers & 4 \\
    Model dimension ($d_{\text{model}}$) & 1024 \\
    Number of heads & 8 \\
    Head dimension ($d_k$) & 128 \\
    Feed-forward dimension ($d_{\text{ff}}$) & 4096 \\
    Activation & GELU \\
    Dropout & 0.1 \\
    Positional embeddings & Learnable, max 6000 \\
    Pooling & Mean \\
    \bottomrule
  \end{tabular}
\end{table}

\subsection{MIDI Encoder --- Full Specification}

\begin{table}[ht]
  \centering
  \caption{MIDI event embedding dimensions.}
  \label{tab:midi_embed}
  \begin{tabular}{lccc}
    \toprule
    Feature & Vocabulary & Embedding dim & Fraction of $d_{\text{model}}$ \\
    \midrule
    Pitch    & 128 & 256 & $d/2$ \\
    Velocity & 128 & 128 & $d/4$ \\
    Duration & 32  & 128 & $d/4$ \\
    \midrule
    \multicolumn{2}{l}{Concatenated} & 512 & $d$ \\
    \multicolumn{2}{l}{+ Linear proj.\ + LayerNorm} & 512 & \\
    \bottomrule
  \end{tabular}
\end{table}

\begin{table}[ht]
  \centering
  \caption{MIDI Transformer encoder specification.}
  \label{tab:midi_transformer}
  \begin{tabular}{lc}
    \toprule
    Hyperparameter & Value \\
    \midrule
    Number of layers & 4 \\
    Model dimension ($d_{\text{model}}$) & 512 \\
    Number of heads & 8 \\
    Head dimension ($d_k$) & 64 \\
    Feed-forward dimension ($d_{\text{ff}}$) & 2048 \\
    Activation & GELU \\
    Dropout & 0.1 \\
    Norm order & Pre-norm \\
    Positional encoding & Sinusoidal, max 2048 \\
    Pooling & Masked mean \\
    Output norm & LayerNorm(512) \\
    \bottomrule
  \end{tabular}
\end{table}

\subsection{Parameter Count Breakdown}

\begin{table}[ht]
  \centering
  \caption{Parameter count breakdown for each Gate~5B model.}
  \label{tab:params}
  \begin{tabular}{lrrrrr}
    \toprule
    Component & D0 & d4a4 & a4r & d4-a4r \\
    \midrule
    Audio CNN             & 3.2M  & 3.2M  & 3.2M  & 3.2M  \\
    Audio Transformer     & 54.5M & 54.5M & 54.5M & 54.5M \\
    Audio pos.\ embeddings & 6.1M  & 6.1M  & 6.1M  & 6.1M  \\
    MIDI embeddings       & 1.2M  & 1.2M  & 1.2M  & 1.2M  \\
    MIDI Transformer      & 10.5M & 10.5M & 10.5M & 10.5M \\
    Audio projection      & 0.8M  & 0.8M  & 0.8M  & 0.8M  \\
    MIDI projection       & 0.4M  & 0.4M  & 0.4M  & 0.4M  \\
    \midrule
    A4 injection          & ---   & 0.01M & ---   & ---   \\
    A4 reverse cross-att  & ---   & ---   & 4.2M  & 4.2M  \\
    D4 injection          & ---   & 0.01M & ---   & 0.01M \\
    A4 concat proj.       & ---   & 0.01M & ---   & ---   \\
    \midrule
    \textbf{Total}        & \textbf{74.2M} & \textbf{75.5M} & \textbf{78.6M} & \textbf{78.9M} \\
    \bottomrule
  \end{tabular}
\end{table}

\section{Gate 4.4: Architecture Family Results}
\label{app:gate44}

\subsection{FiLM Conditioning}

Feature-wise Linear Modulation~\citep{perez2018film} modulates Transformer layer activations:
\begin{equation}
  F'_\ell = (1 + \gamma_\ell) \odot F_\ell + \beta_\ell,
\end{equation}
where $\gamma_\ell, \beta_\ell \in \R^{d_{\text{model}}}$ are generated from the descriptor via a 2-layer MLP:
\begin{equation}
  [\gamma_\ell, \beta_\ell] = W_2\,\ReLU(W_1\,\bar{d} + b_1) + b_2,
\end{equation}
with $\bar{d} = \text{MeanPool}(d)$ being the temporally averaged descriptor.

Three FiLM variants were tested: \texttt{film-a4} (A4 only), \texttt{film-d4} (D4 only), \texttt{film-dual} (both). None improved over D0 (best: 59.4\%, D0: 60.2\%).

\subsection{Mixture-of-Experts}

We implemented a sparsely-gated MoE layer~\citep{shazeer2017moe} replacing the feed-forward sub-layer in the final Transformer block:
\begin{equation}
  \text{MoE}(x) = \sum_{i=1}^{E} g_i(x) \cdot \text{FFN}_i(x), \quad g(x) = \text{TopK}\!\big(\softmax(W_g\,x),\; k\big),
\end{equation}
with $E = 4$ or $8$ experts and $k = 2$. A load-balance loss $\mathcal{L}_{\text{bal}}$ and entropy regularizer $\mathcal{L}_{\text{ent}}$ encourage uniform expert utilization.

Five MoE variants were tested (\texttt{moe-a4}, \texttt{moe-dual}, \texttt{moe-a4-v2} through \texttt{v4}) with varying numbers of experts, top-$k$ values, and gating architectures. Best: 60.2\% (moe-a4-v2), matching D0 exactly.

\subsection{Third Tower}

The Third Tower architecture introduces a separate descriptor encoder whose representations are aligned with both audio and MIDI via auxiliary VICReg losses:
\begin{equation}
  \mathcal{L}_{\text{t3}} = \mathcal{L}_{\text{VICReg}}(z_a, z_m) + \alpha\,\mathcal{L}_{\text{VICReg}}(z_a, z_d) + \beta\,\mathcal{L}_{\text{VICReg}}(z_m, z_d),
\end{equation}
where $z_d$ is the descriptor tower output.

Three variants were tested:
\begin{itemize}[nosep]
  \item \texttt{t3-wt} (weighted loss, $\alpha = \beta = 0.5$): $S = 67.6\%$ ($+7.4\pp$). Best of the family.
  \item \texttt{t3-tri} (triplet-style): $S = 65.0\%$ ($+4.8\pp$).
  \item \texttt{t3-anc} (anchor, descriptor as anchor for both): $S = 42.2\%$ ($-18.0\pp$). Catastrophic.
\end{itemize}

The t3-wt variant was trained to 50 epochs, achieving $S = 81.2\%$---competitive with direct injection methods.

\section{Training Hyperparameters}
\label{app:hyperparams}

\begin{table}[ht]
  \centering
  \caption{Complete training hyperparameters for Gate~5B canonical models.}
  \label{tab:hyperparams}
  \begin{tabular}{lcccc}
    \toprule
    Parameter & D0 & d4a4 & a4r & d4-a4r \\
    \midrule
    Optimizer & \multicolumn{4}{c}{AdamW ($\beta_1{=}0.9$, $\beta_2{=}0.999$)} \\
    Weight decay & \multicolumn{4}{c}{0.01} \\
    Batch size & \multicolumn{4}{c}{16} \\
    \midrule
    Total epochs & 60 & 60 & 30 & 30 \\
    Best epoch & 50 & 50 & 29 & 30 \\
    Warmup steps & \multicolumn{4}{c}{500} \\
    \midrule
    Schedule & Cosine-tail & Cosine-60 & Cosine-30 & Cosine-30 \\
    LR floor & 0.10 & --- & --- & --- \\
    LR tail end & 0.02 & --- & --- & --- \\
    Cosine ref epochs & 30 & 60 & 30 & 30 \\
    \midrule
    VICReg $\lambda_{\text{inv}}$ & \multicolumn{4}{c}{10} \\
    VICReg $\lambda_{\text{var}}$ & \multicolumn{4}{c}{10} \\
    VICReg $\lambda_{\text{cov}}$ & \multicolumn{4}{c}{1} \\
    \midrule
    Segment length & \multicolumn{4}{c}{4\,s (96,000 samples at 24\,kHz)} \\
    Segment hop & \multicolumn{4}{c}{1\,s} \\
    Max MIDI notes & \multicolumn{4}{c}{2048} \\
    Duration buckets & \multicolumn{4}{c}{32 (log-spaced, 0.05--4\,s)} \\
    \midrule
    DataLoader workers & \multicolumn{4}{c}{8} \\
    Pin memory & \multicolumn{4}{c}{True} \\
    Prefetch factor & \multicolumn{4}{c}{2} \\
    \midrule
    Data augmentation & \multicolumn{4}{c}{None (deliberate)} \\
    \bottomrule
  \end{tabular}
\end{table}

\paragraph{No data augmentation.} We deliberately omit data augmentation (pitch shift, time stretch, noise injection) to isolate the effect of descriptor injection from augmentation-based regularization. This conservative choice means our results represent a lower bound on achievable performance.

\paragraph{Scheduler variants explored.}
\begin{itemize}[nosep]
  \item \textbf{Cosine}: Standard cosine annealing~\citep{loshchilov2017sgdr} from base LR to 0 over $T_{\text{ref}}$ epochs.
  \item \textbf{Cosine-tail}: Cosine annealing with a floor ($10\%$ base LR) and linear tail ($\to 2\%$) for epochs beyond $T_{\text{ref}}$.
  \item \textbf{Trapezoidal}: Linear warmup $\to$ constant hold $\to$ linear decay.
\end{itemize}

\paragraph{Evaluation configuration (canonical, pinned).}
\begin{itemize}[nosep]
  \item Pool size: 256 random items
  \item Number of queries: 500
  \item Hard negatives: 64 (same piece, different time)
  \item Semi-hard negatives: 32 (same composer, different piece)
  \item Random seed: 42
  \item Embedding dimension: 256
\end{itemize}

\section{Additional Visualizations}
\label{app:visualizations}

\subsection{t-SNE and UMAP Embedding Spaces}

\cref{fig:tsne_appendix,fig:umap_appendix} show t-SNE and UMAP projections of 2000 audio (blue) and MIDI (red) embeddings for each model. Matched pairs are connected by light gray lines.

\newcommand{\gatevizdir}{figures}
\newcommand{\vizpanel}[2]{%
  \IfFileExists{#1}{%
    \includegraphics[width=\linewidth]{#1}%
  }{%
    \fbox{\parbox{\linewidth}{\centering\vspace{2.4cm}\small #2\\(missing file)\vspace{2.4cm}}}%
  }%
}

\begin{figure}[ht]
  \centering
  \begin{subfigure}[b]{0.48\textwidth}
    \vizpanel{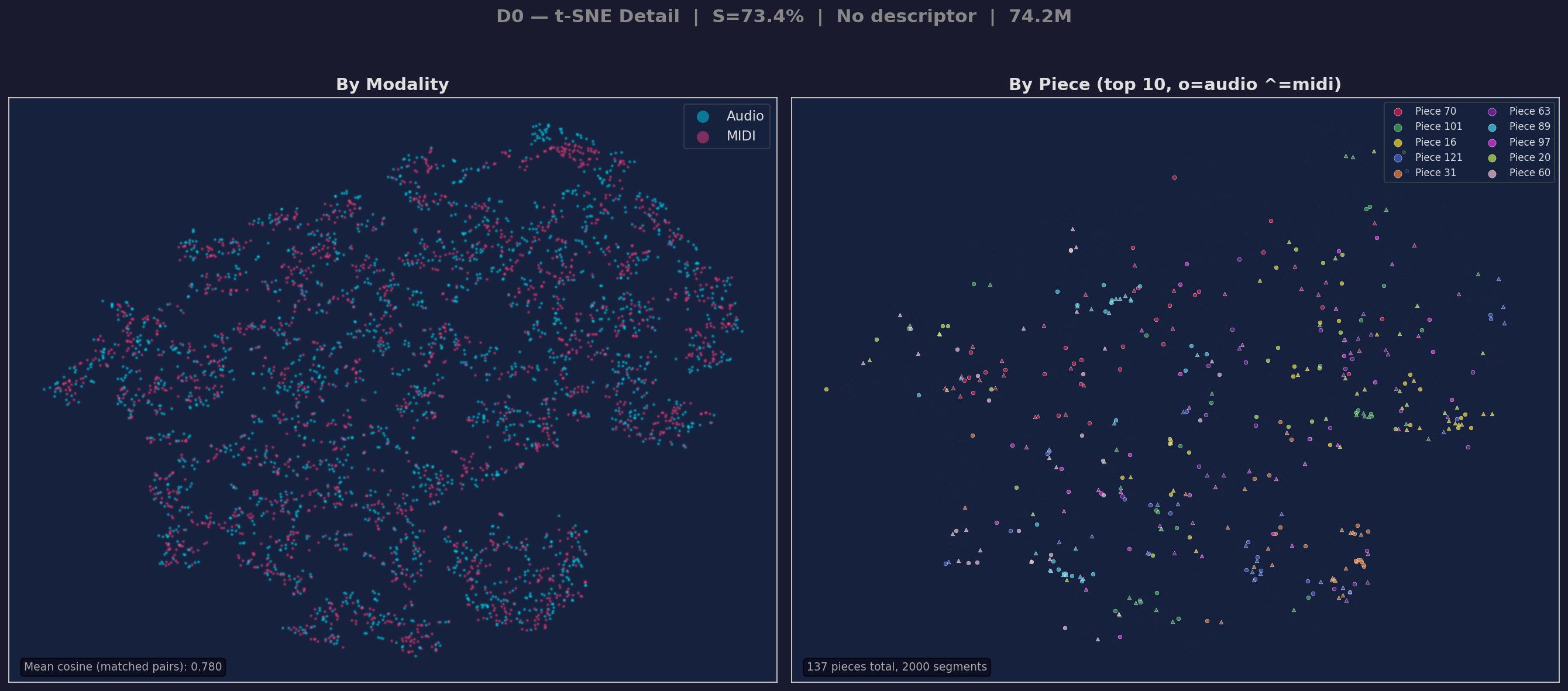}{D0 t-SNE}
    \caption{D0 --- t-SNE}
  \end{subfigure}
  \hfill
  \begin{subfigure}[b]{0.48\textwidth}
    \vizpanel{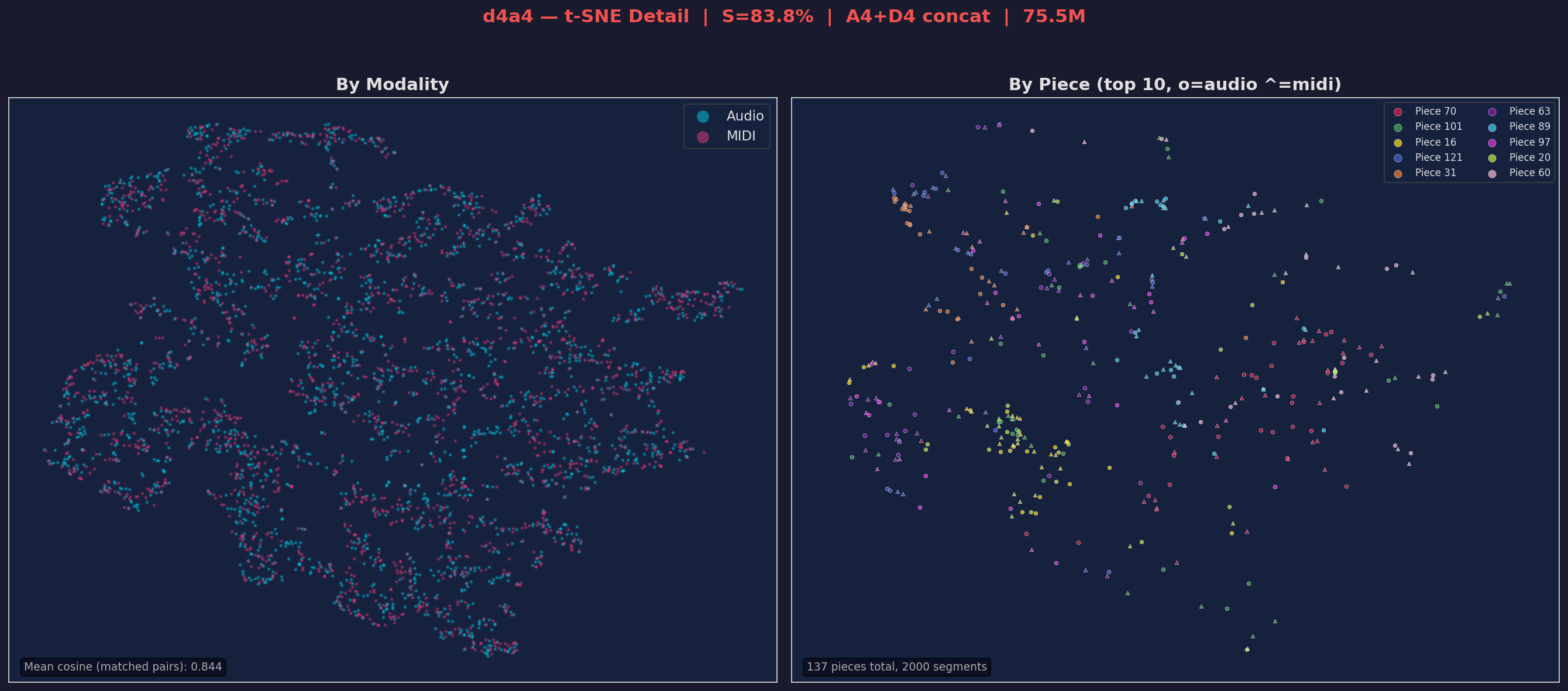}{d4a4 t-SNE}
    \caption{d4a4 --- t-SNE}
  \end{subfigure}

  \vspace{0.5cm}

  \begin{subfigure}[b]{0.48\textwidth}
    \vizpanel{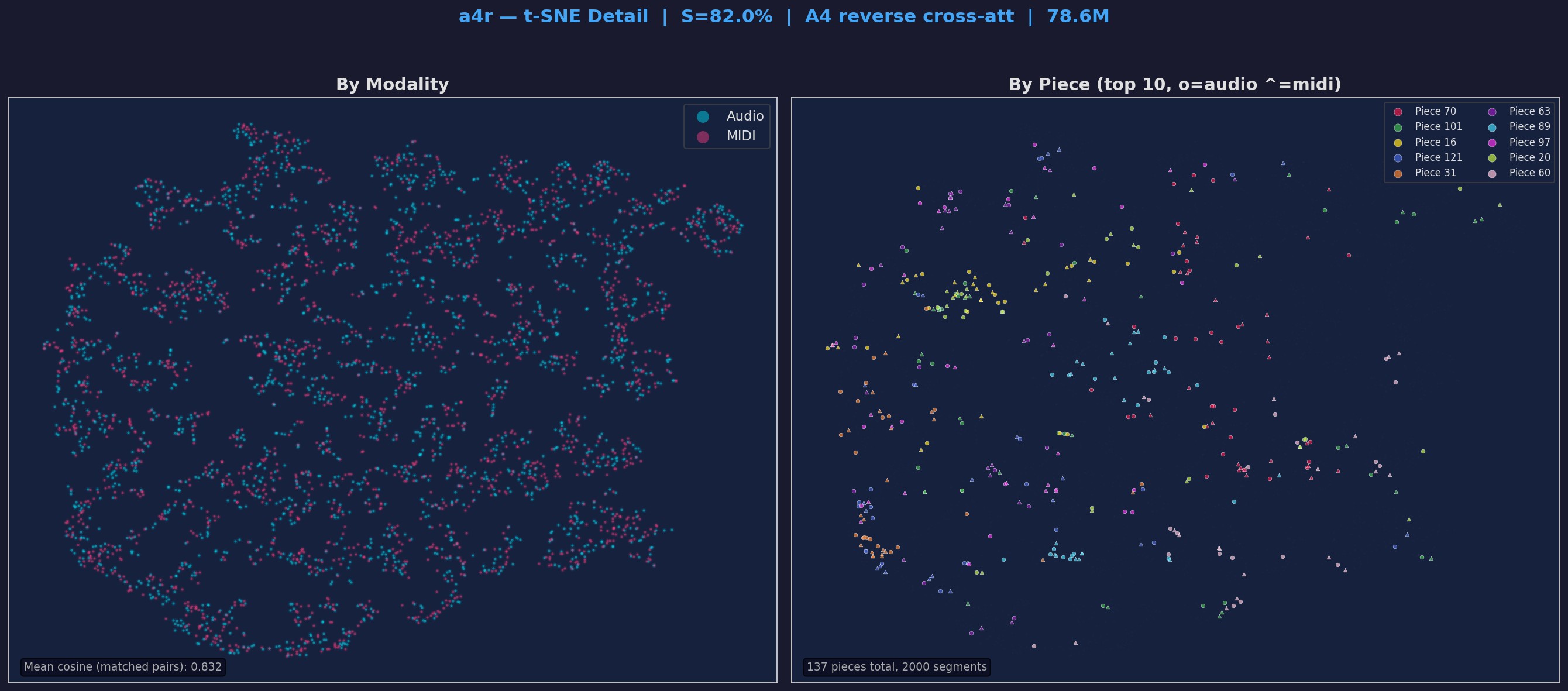}{a4r t-SNE}
    \caption{a4r --- t-SNE}
  \end{subfigure}
  \hfill
  \begin{subfigure}[b]{0.48\textwidth}
    \vizpanel{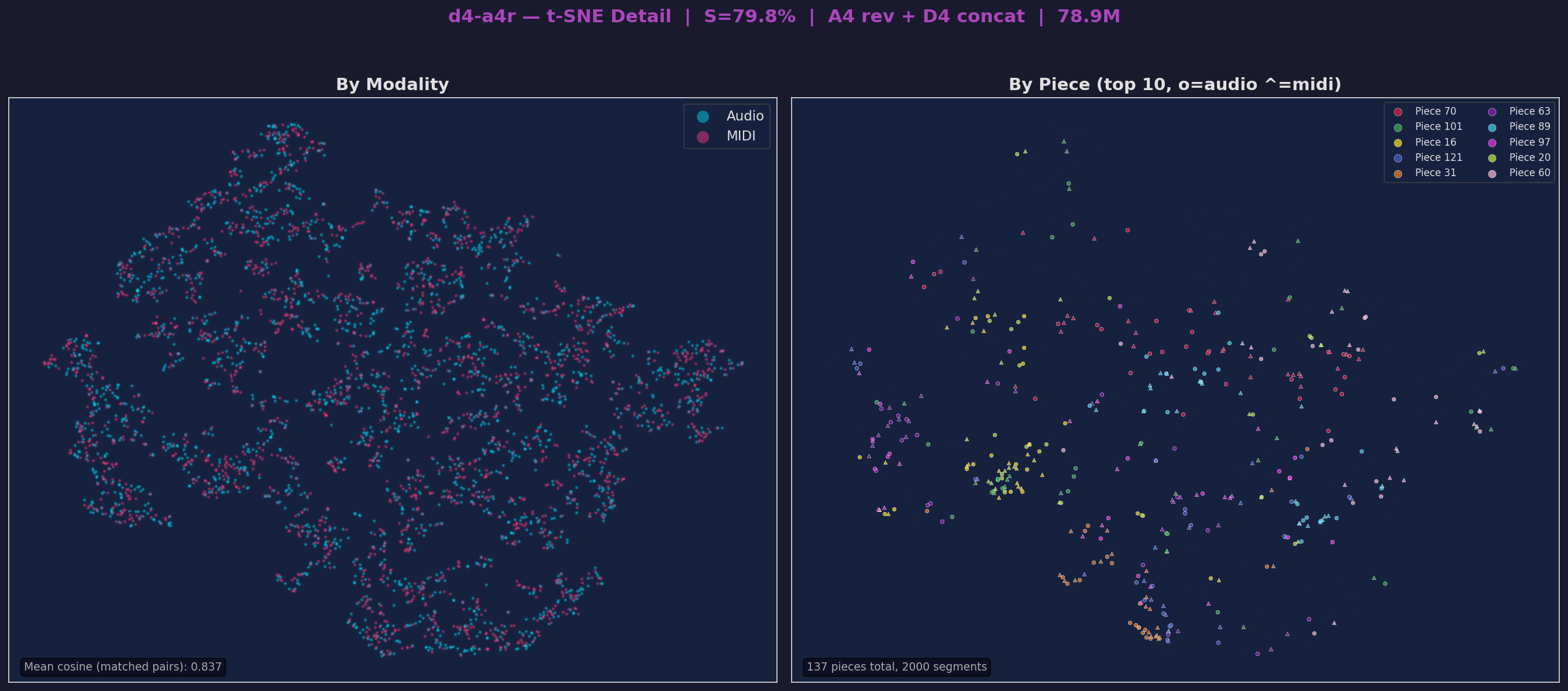}{d4-a4r t-SNE}
    \caption{d4-a4r --- t-SNE}
  \end{subfigure}
  \caption{t-SNE embeddings for the 4 canonical models from the seed-42 checkpoint. Blue: audio; red: MIDI. Lines connect matched pairs. Figures are rendered from Test~10 exported PNGs.}
  \label{fig:tsne_appendix}
\end{figure}

\begin{figure}[ht]
  \centering
  \begin{subfigure}[b]{0.48\textwidth}
    \vizpanel{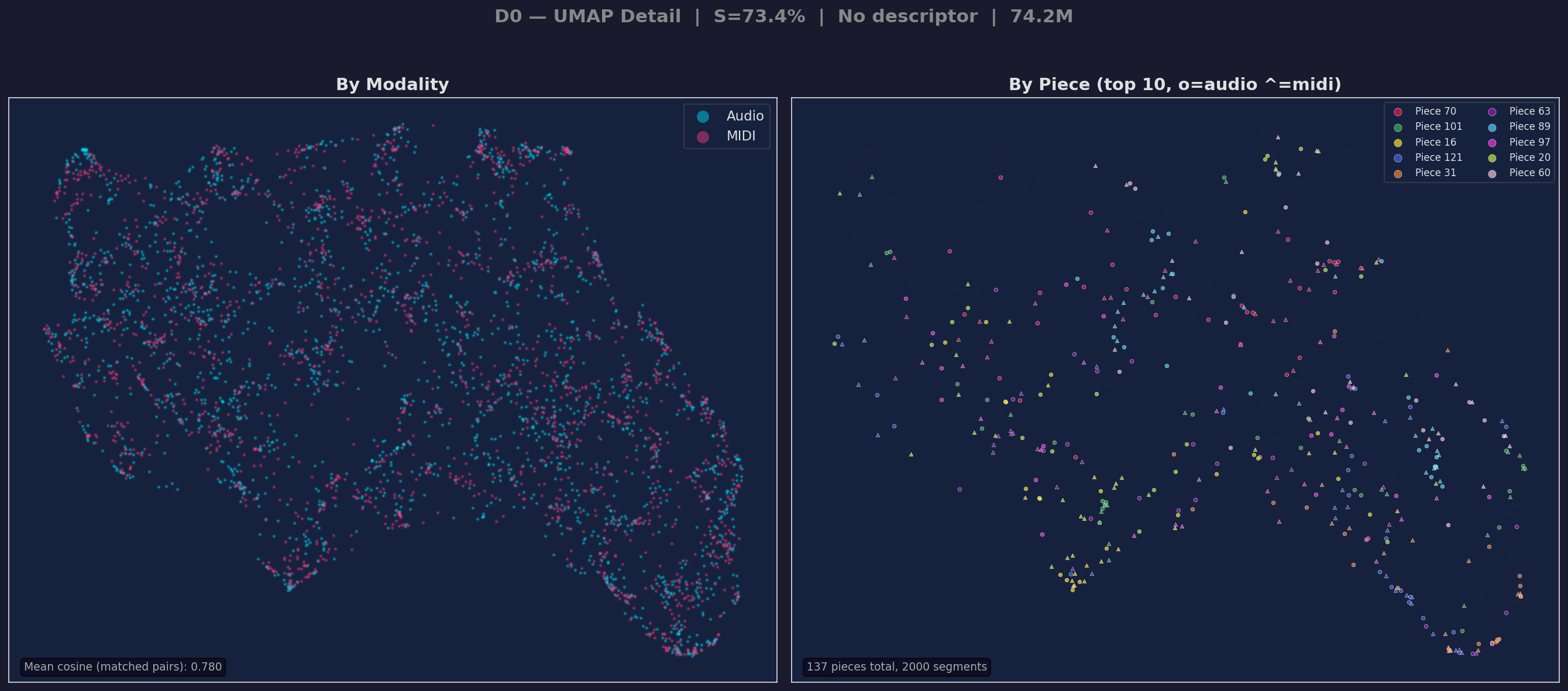}{D0 UMAP}
    \caption{D0 --- UMAP}
  \end{subfigure}
  \hfill
  \begin{subfigure}[b]{0.48\textwidth}
    \vizpanel{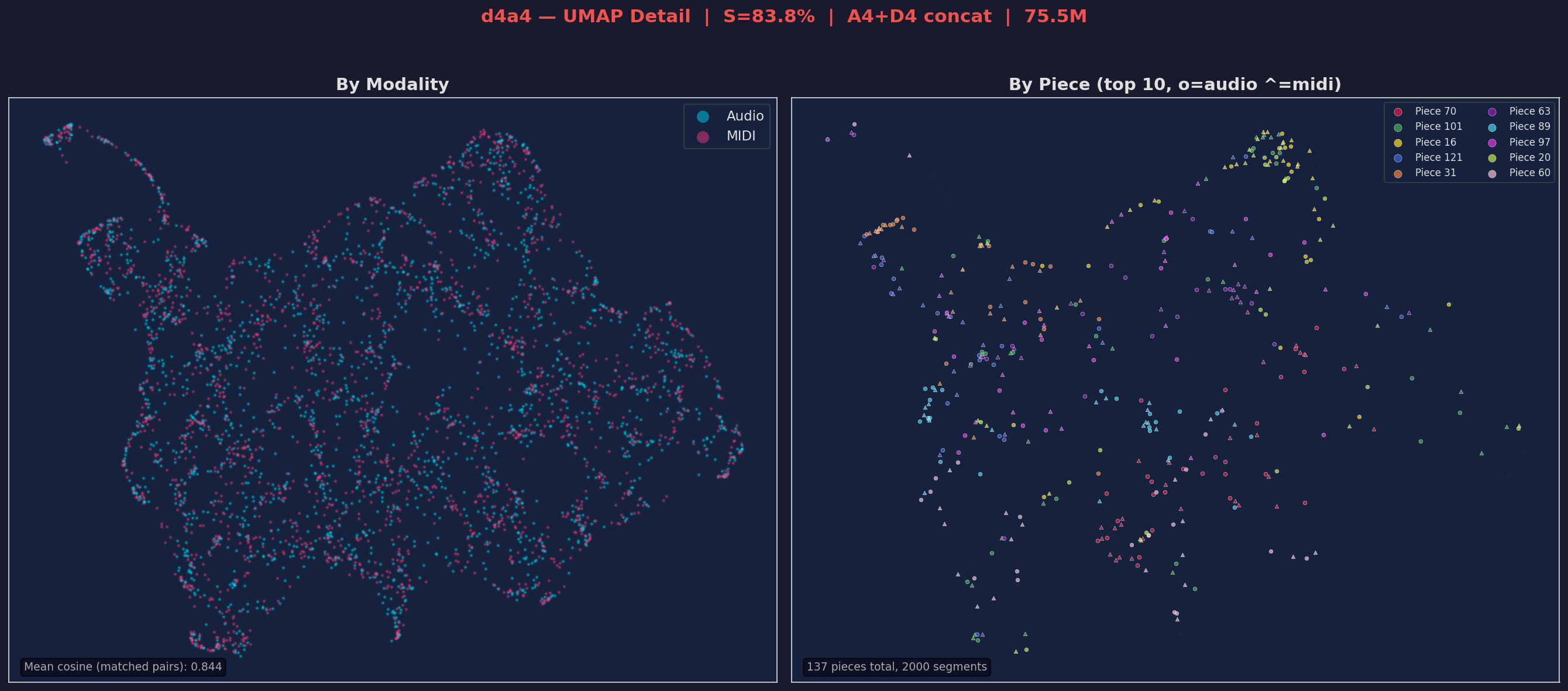}{d4a4 UMAP}
    \caption{d4a4 --- UMAP}
  \end{subfigure}

  \vspace{0.5cm}

  \begin{subfigure}[b]{0.48\textwidth}
    \vizpanel{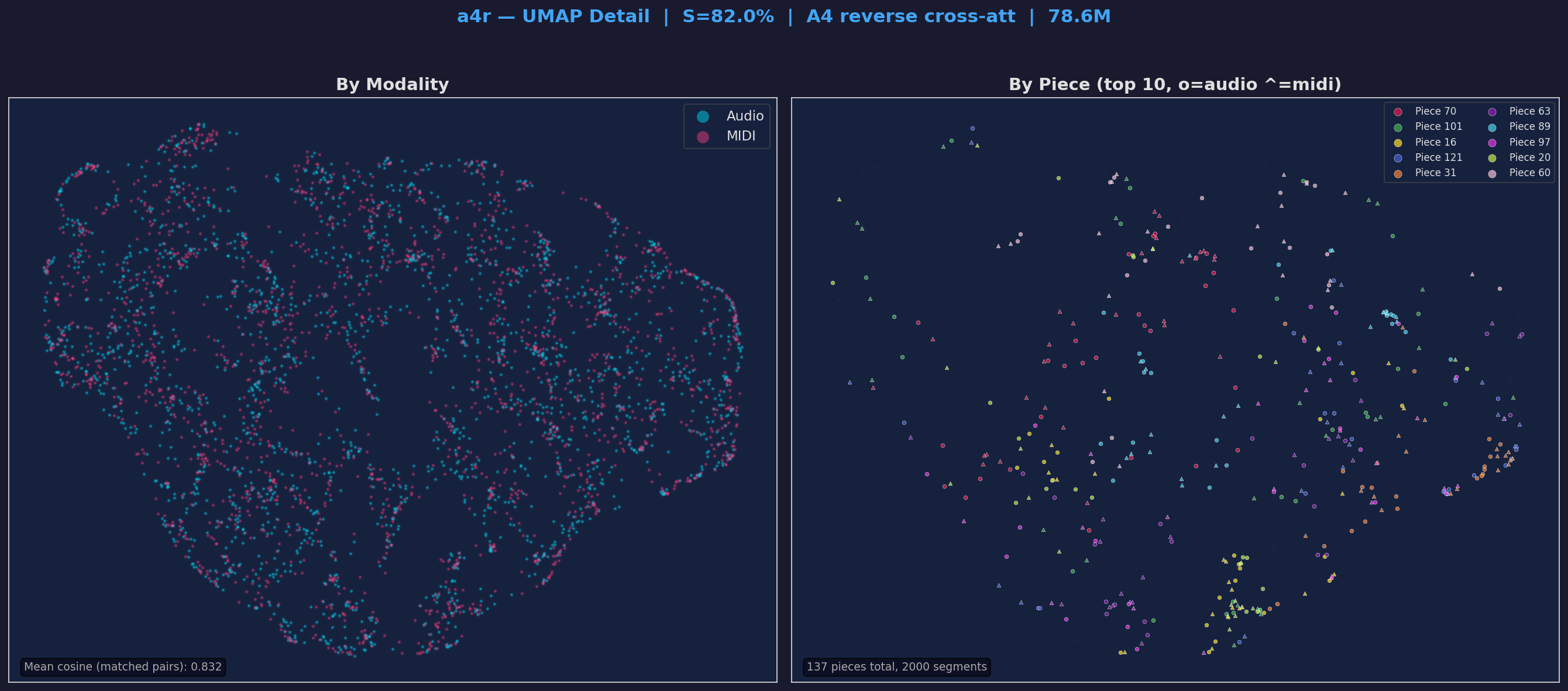}{a4r UMAP}
    \caption{a4r --- UMAP}
  \end{subfigure}
  \hfill
  \begin{subfigure}[b]{0.48\textwidth}
    \vizpanel{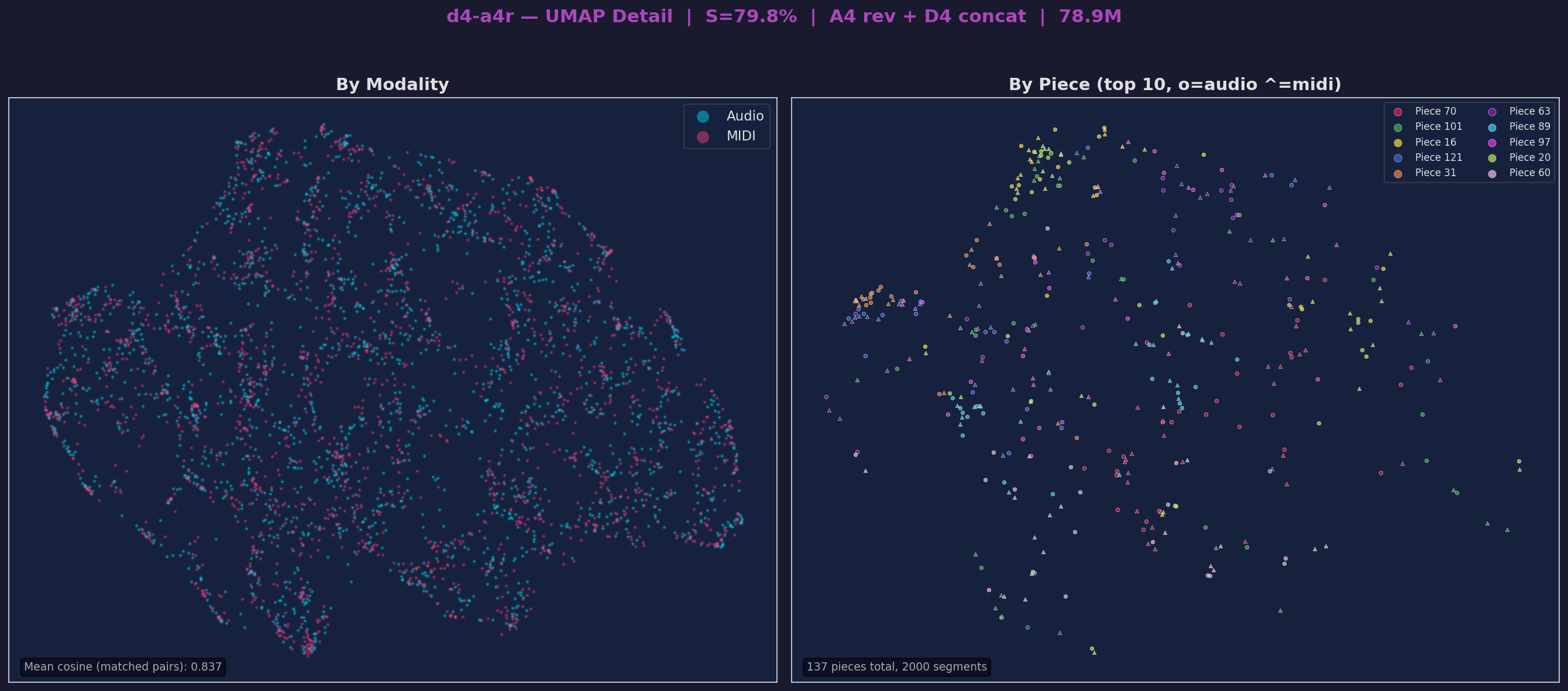}{d4-a4r UMAP}
    \caption{d4-a4r --- UMAP}
  \end{subfigure}
  \caption{UMAP embeddings for the 4 canonical models from the seed-42 checkpoint (same split and sample count as t-SNE).}
  \label{fig:umap_appendix}
\end{figure}

\subsection{Alignment Cosine Distributions}

The distribution of cosine similarities between matched audio--MIDI embedding pairs provides a complementary view to the retrieval metrics. Higher mean cosine with lower variance indicates more consistent alignment.

\begin{figure}[ht]
  \centering
  \vizpanel{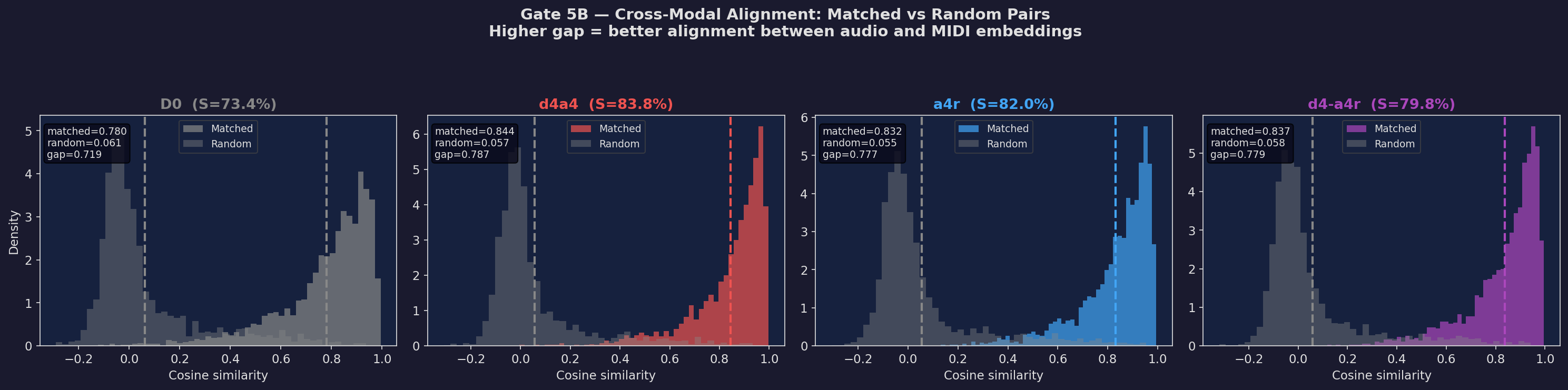}{Alignment cosine distribution}
  \caption{Distribution of cosine similarity for matched audio--MIDI pairs across canonical models from the seed-42 checkpoint (Test~10).}
  \label{fig:cosine_align_appendix}
\end{figure}